\documentclass[12pt]{article}
\usepackage[hyperfootnotes=false]{hyperref}
\usepackage{epsfig}
\usepackage{amsmath}
\usepackage{amssymb}
\usepackage{caption}

\numberwithin{equation}{section}

\usepackage{setspace}
\usepackage{graphicx}
\usepackage{color}

\def\be{\begin{equation}}
\def\ee{\end{equation}}
\def\ba#1\ea{\begin{align}#1\end{align}}
\def\bg#1\eg{\begin{gather}#1\end{gather}}
\def\bm#1\em{\begin{multline}#1\end{multline}}
\def\bmd#1\emd{\begin{multlined}#1\end{multlined}}

\def\d{\delta}

\def\e{\epsilon}

\def\l{\lambda}

\def\m{\mu}
\def\n{\nu}
\def\p{\phi}

\def\r{\rho}
\def\s{\sigma}

\def\t{\tau}

\def\w{\omega}

\def\y{\psi}

\def\la{\label}

\def\re{\ref}
\def\er{\eqref}
\def\se{\section}
\def\sse{\subsection}

\def\fr{\frac}
\def\na{\nabla}
\def\pa{\partial}

\def\wtd{\widetilde}
\def\ol{\overline}

\def\eq{\equiv}

\def\cd{\cdots}
\def\ap{\approx}
\def\mr{\mathring}

\def\qu{\quad}
\def\qqu{\qquad}

\def\lt{\left}
\def\rt{\right}
\def\({\left(}
\def\){\right)}
\def\[{\left[}
\def\]{\right]}
\def\<{\langle}
\def\>{\rangle}
\def\lra{\leftrightarrow}

\def\Tr{\operatorname{Tr}}
\def\bZ{{\mathbb Z}}

\def\zb{\bar{z}}

\def\ext{\operatorname*{ext}}

\def\Wald{\text{Wald}}
\def\anomaly{\text{anomaly}}
\def\extrinsic{\text{extrinsic}}
\def\gen{\text{gen}}
\def\rel{\text{rel}}
\def\bulk{\text{bulk}}
\def\diff{\text{diff}}

\def\nref#1{(\ref{#1})}

\begin{document}

\begin{titlepage}
  \bigskip

  \bigskip\bigskip\bigskip\bigskip

  \bigskip

\centerline{\Large \bf {Entropy, Extremality, Euclidean Variations,}}\bigskip \centerline{\Large \bf {and the Equations of Motion}}

  \bigskip

  \begin{center}

\bf {Xi Dong$^{1,2}$, Aitor Lewkowycz$^3$}

  \bigskip \rm
  \bigskip

$^1${\it School of Natural Sciences, Institute for Advanced Study, Princeton, NJ 08540, USA}

  \smallskip
 
$^2${\it Department of Physics, University of California, Santa Barbara, CA 93106, USA}

  \smallskip
 
$^3${\it  Stanford Institute for Theoretical Physics, Department of Physics,\\ Stanford University, Stanford, CA 94305, USA}

  \smallskip 
 
  \vspace{1cm}
  \end{center}

  \bigskip\bigskip

  \bigskip\bigskip

\begin{abstract}
We study the Euclidean gravitational path integral computing the R\'{e}nyi entropy and analyze its behavior under small variations.  We argue that, in Einstein gravity, the extremality condition can be understood from the variational principle at the level of the action, without having to solve explicitly the equations of motion.  This set-up is then generalized to arbitrary theories of gravity, where we show that the respective entanglement entropy functional needs to be extremized. We also extend this result to all orders in Newton's constant $G_N$, providing a derivation of quantum extremality. Understanding quantum extremality for mixtures of states provides a generalization of the dual of the boundary modular Hamiltonian which is given by the bulk modular Hamiltonian plus the area operator, evaluated on the so-called modular extremal surface. This gives a bulk prescription for computing the relative entropies to all orders in $G_N$. We also comment on how these ideas can be used to derive an integrated version of the equations of motion, linearized around arbitrary states.
\end{abstract}

\end{titlepage}

\tableofcontents

%%%%%%%%%%%%%%%%%%%%%%%%%%%%%%%%%%%%%%%%%%%%%%%%%%
\section{Introduction and summary of results}

Quantum entanglement has become a crucial aspect of understanding many physical systems including quantum gravity. A universal property of quantum gravity is that entropy satisfies an area law.  This was first discovered for black holes \cite{Bekenstein:1973ur, Bardeen:1973gs, Hawking:1974sw}, and more recently it was generalized in the context of AdS/CFT correspondence \cite{Maldacena:1997re, Gubser:1998bc, Witten:1998qj} by Ryu and Takayanagi \cite{Ryu:2006bv, Ryu:2006ef}.  They gave an elegant prescription for the entanglement entropy of any spatial region $R$ in a holographic boundary theory in terms of the area of an extremal surface in the bulk spacetime:
\be\la{rt}
S_R = \ext_{X \sim R} \fr{A(X)}{4G_N}.
\ee
Here the entanglement entropy is defined in the boundary theory as the von Neumann entropy $S_R\eq -\Tr \r_R \log \r_R$ of the reduced density matrix $\r_R$, and is a measure of entanglement between the region $R$ and its complement $\ol{R}$.  The constraint $X \sim R$ means that the Ryu--Takayanagi (RT) surface $X$ is homologous to the boundary region $R$, and $G_N$ denotes Newton's constant.  This prescription for holographic entanglement entropy was derived from AdS/CFT in \cite{Lewkowycz:2013nqa}.  Furthermore, it is valid in general time-dependent cases \cite{Hubeny:2007xt,Dong:2016hjy}.

In general, the gravitational theory in the bulk is described at low energies in terms of Einstein gravity corrected by higher derivative interactions.  These interactions generate higher derivative corrections to the RT formula \er{rt}.  A prescription for these corrections was given in \cite{Dong:2013qoa,Camps:2013zua} and has the form
\be
A_{\gen} = S_{\Wald} + S_{\extrinsic}
\ee
where the first term is the Wald entropy and the second consists of corrections from the extrinsic curvature of the RT surface. Since $A_{\gen}$ is the full classical contribution to the gravitational entropy, we will refer to it as the ``generalized area''.\footnote{For Einstein gravity, $A_{\gen}=\frac{A}{4 G_N}$.} However, it has been an open question whether the extremization procedure in \er{rt} works for general higher derivative gravity, using variations of the action.  Our first result is that it does:
\be\la{rth}
S_R = \ext_{X \sim R} A_{\gen}(X).
\ee

As a byproduct of this result, one can generalize the derivation of the integrated linearized equations of motion from the first law of entanglement \cite{Blanco:2013joa,Wong:2013gua,Lashkari:2013koa,Faulkner:2013ica} to arbitrary regions and states. This is done by defining the variation of the modular Hamiltonian using the replica trick and from the linearized equations of motion for an arbitrary state one should in principle be able to get the nonlinear equations of motion. 

The RT prescription \er{rt} and its higher derivative generalization \er{rth} are valid in the large-$N$ limit of the boundary theory.  Beyond the leading order in this limit, they would receive $1/N$ corrections from quantum effects in the bulk.  A natural prescription for these quantum corrections is
\be\la{rtq}
S_R = \ext_{X \sim R} S_{\gen}(X),\qu
S_{\gen} \eq \langle A_{\gen} \rangle+ S_{\bulk},
\ee
where the ``generalized entropy'' $S_{\gen}$ is the sum of the expectation value of the generalized area $\langle A_{\gen} \rangle$ and a bulk entanglement entropy $S_{\bulk}$.  The bulk entanglement entropy is defined with respect to the bulk spatial region between the RT surface $X$ and the boundary region $R$.  The domain of dependence of this region defines the notation of the entanglement wedge \cite{Czech:2012bh,Wall:2012uf,Headrick:2014cta}.  It is worth noting that after extremization $X$ is known as the quantum extremal surface.

The prescription \er{rtq} agrees with the one-loop result of \cite{Faulkner:2013ana,Barrella:2013wja} and was conjectured in \cite{Engelhardt:2014gca} to hold for all loops.  Our second result is to establish this from AdS/CFT to all orders in $1/N$.

Furthermore, entanglement entropy is not the only measure of quantum entanglement.  To better understand the structure of entanglement, we also need the modular Hamiltonian
\be
K_{\r} \eq -\log\r
\ee
for a quantum state described by the density matrix $\r$, as well as the relative entropy
\be
S_{\rel}(\r|\s) \eq \Tr \[\r\log\r - \r\log\s\]
\ee
which is a measure of distinguishability between an arbitrary state $\r$ and a reference state $\s$.  Our third result is 
\be\la{rtm}
\langle K_{R,\sigma} \rangle_{\rho} = \ext_{X \sim R} \[\langle A_{\gen}^{X} \rangle_{\rho} +\langle K_{\bulk,\sigma}^{X} \rangle_{\rho}\]
\ee
where $K_{R,\sigma}$ is the modular Hamiltonian for the boundary region $R$ for the state $\s$, $A_{\gen}$ is viewed as an operator on the surface $X$ giving its generalized area, and $K_{\bulk}$ is the bulk modular Hamiltonian in the spatial region between $X$ and $R$.  After extremization we call $X$ the ``modular extremal surface'' for the state $\s$.

Using the prescription \er{rtm} for the modular Hamiltonian, we find for the relative entropy
\be\la{rem}
S_{\rel}(\r|\s) = \<A_{\gen}^{X_\s} +K_{\bulk,\s}^{X_\s}\>_\r - \<A_{\gen}^{X_\r} +K_{\bulk,\r}^{X_\r}\>_\r
\ee
where $X_\s$ and $X_\r$ are modular extremal surfaces defined by \er{rtm} for the states $\s$ and $\r$ respectively.  Here we have dropped explicit references to the boundary region $R$ for brevity, and $\<\cd\>_\r$ denotes the expectation value $\Tr\left(\rho \cd \right)$ in the state $\r$.

The results \er{rtm} and \er{rem} agree with one-loop results of \cite{Jafferis:2015del}.  As we will show using AdS/CFT, they are valid to all orders in $1/N$.  It is interesting to note from \er{rem} that the boundary relative entropy is equal to the bulk relative entropy only at the one-loop order \cite{Jafferis:2015del}, and they generally differ at two loops or higher.  This is because the two modular extremal surfaces $X_\s$ and $X_\r$ differ by $O(G_N)$ in general.

Recently, the AdS/CFT dictionary has been clarified by viewing holography as a quantum error correcting code \cite{Almheiri:2014lwa}.  The relation between the bulk and boundary relative entropy was used in \cite{Dong:2016eik} to prove a theorem for reconstructing bulk operators in the entanglement wedge of $R$ in terms of boundary operators on $R$, and the one-loop result can be used to obtain an explicit large-$N$ reconstruction formula in terms of the modular flow \cite{Faulkner:2017vdd}.  As we will see, the all-loop result \er{rem} can be used to extend the reconstruction theorem to all orders in $1/N$, at least for bulk operators at a fixed distance away from the RT surface, but it is not yet clear how to generalize the modular flow construction beyond one loop.  A related issue is that the complementary recovery property discussed in \cite{Harlow:2016vwg} holds only at the one-loop order.

The outline of this paper is as follows.  We begin in Section \re{secce} with a review of the classical statement of extremality and rephrase it in a way that can easily be generalized to arbitrary theories of gravity, using variations of the action. Section \re{EOM} is independent of the rest of the paper and uses the variational principle to derive the integrated equations of motion around an arbitrary background using the first law. In Section \re{qext}, we generalize the classical discussion of Section \re{secce} by including quantum fields in the bulk theory, providing a derivation of quantum extremality. In Section \re{mext}, we use quantum extremality for mixtures of states to write a formula for the bulk dual of the modular Hamiltonian to all orders in $G_N$. We conclude with some closing thoughts in the discussion.  

%%%%%%%%%%%%%%%%%%%%%%%%%%%%%%%%%%%%%%%%%%%%%%%%%%
\section{Classical statement of extremality from variations}\la{secce}

Let us start with a review of the replica trick applied to AdS/CFT.  In the boundary theory, the von Neumann entropy may be determined by the $n\to1$ limit of the R\'{e}nyi entropy
\be
S_n \eq \fr{1}{1-n} \log \Tr \r^n,
\ee
where $n$ is known as the R\'{e}nyi index.  When $n$ is an integer greater than $1$, the R\'{e}nyi entropy can be calculated from
\be
S_n = \fr{1}{1-n} \log \fr{Z_n}{Z_1^n},
\ee
where $Z_n$ is the partition function of the boundary theory on a manifold known as the $n$-fold branched cover.  This partition function can be calculated via AdS/CFT.  In the large-$N$ limit, we find the solution $M_n$ to the bulk equations of motion with the $n$-fold cover as the boundary condition and calculate its on-shell action $I_n$.  Up to $1/N$ corrections, we have $\log Z_n = -I_n$.  When there are more than one bulk solution, we choose the dominant one which has the smallest on-shell action.

The $n$-fold cover on the boundary enjoys a $\bZ_n$ symmetry permuting the $n$ replicas cyclically.  As in \cite{Lewkowycz:2013nqa}, we assume that the $\bZ_n$ replica symmetry extends to the dominant bulk solution $M_n$. Let us take the quotient of the bulk solution $M_n$ by the $\bZ_n$ replica symmetry.
This quotient amounts to considering the action  $\hat{I}_n=I_n/n$ which can be thought of as the on-shell action of the orbifold geometry $\hat{M}_n \eq M_n /\bZ_n$.  The orbifold has a conical singularity at the $\mathbb{Z}_n$ fixed points.  The derivative of the orbifold action with respect to $n$ is the modular entropy introduced in \cite{Dong:2016fnf}:
\begin{equation}\la{stilde}
\tilde{S}_n\equiv -n^2 \partial_n \(\frac{1}{n} \log \Tr \r^n\) = n^2 \partial_n \hat{I}_n.
\end{equation}
Since the orbifold geometry is seemingly singular, when doing variations one has to be careful with possible boundary terms at the tip of the cone.  In other words, \er{stilde} reduces to a boundary term on the conical defect, and taking the $n\to1$ limit we find the von Neumann entropy $S$ in terms of some geometric quantity $A_{\gen}$ on a codimension-2 surface $X$.

The goal of this section is to show that for classical theories of gravity, the equations of motion close to $n\ap1$ imply that the surface $X$ has to be extremal with respect to the entanglement entropy functional $A_{\gen}$:
\begin{equation}
\delta_{\diff} A_{\gen}=0
\end{equation}
where $\diff$ denotes to a diffeomorphism that would change the location of $X$ where the functional is evaluated. 

%%%%%%%%%%%%%%%%%%%%%%%%%%%%%%%%%%%%%%%%%%%%%%%%%%
\subsection{Double variations}

If we vary the action around the solution $g_n$ to the equations of motion with an off-shell deformation $\d g_n$ that preserves the conical deficit angle and vanishes on the asymptotic boundary, we have
\be\la{dhi}
\delta \hat{I}_n=\int_{\hat{M}_n} E_n \d g_n +\lt.\int_{\partial \hat{M}_n} \Theta(g_n,\d g_n) \rt|_{r=\epsilon} =0
\ee
where we have used the notation of \cite{Iyer:1994ys}: $E_n \equiv \delta \hat I_n / \delta g$ denotes the equations of motion at integer $n$, and $\Theta(g_n,\d g_n)$ is the boundary term at the tip of the cone, linear in $\d g$ and obtained from integrating the Lagrangian by parts after a variation.  The solution $g_n$ satisfies the equations of motion, leading to $E_n=0$.  The boundary term is evaluated on a regulated surface $r=\e$ where $r$ is the radial distance from the tip of the cone, and we take the $\e\to0$ limit at the end of the calculation.  The claim of \er{dhi} is that the boundary term vanishes in this limit.

For integer $n$, it is clear that \er{dhi} holds, since we can go to the parent space $M_n$ where there is no physical boundary at the $\bZ_n$ fixed points.

In the next subsection, we will argue that \er{dhi} holds for general values of $n$. For now we will explore the consequences of this, saving the details for later. Since \er{dhi} is zero for any $n$, its derivative with respect to $n$ is also zero:
\begin{equation}\la{pndi}
\partial_n \delta \hat{I}_n \big|_{n=1}=0.
\end{equation}
Note that this follows as long as the equations of motion are obeyed at $n \ap 1$.  

We can take the two variations $\partial_n$ and $\delta$ in \er{pndi} in the opposite order, so that $\partial_n$ gives us the entanglement entropy functional $A_{\gen}$ for a metric in the neighbourhood of the on-shell metric. Up until now we have kept the variation of the metric $\d g_n$ arbitrary except for the boundary conditions of preserving the conical deficit angle and vanishing on the asymptotic boundary.  Let us now choose $\d g_n$ to become a diffeomorphism at $n=1$.\footnote{We do not put additional constraints on $\d g_n$ away from $n=1$ except for the boundary conditions.  In general $\d g_n$ will be off-shell at finite $n-1$ because the conical boundary condition essentially fixes the on-shell solution as $g_n$.}  If we consider the variations in the opposite order for a diffeomorphism at $n=1$, we obtain
\ba
\partial_n \hat{I}_n|_{n=1} & = \lim_{\delta n\rightarrow 0} \frac{\hat{I}_{1+\delta n}[g_{1+\delta n}]-\hat{I}_{1}[g_{1}]}{\delta n}= A_{\gen} \nonumber \\
\lt.\(\partial_n  \hat{I}_n+\delta \partial_n  \hat{I}_n\)\rt|_{n=1} &=\lim_{\delta n\rightarrow 0} \frac{\hat{I}_{1+\delta n}[g_{1+\delta n}+\delta g_{1+\delta n}]-\hat{I}_{1}[g_{1}]}{\delta n}=A_{\gen}+ \delta_{\diff} A_{\gen} \label{variationn}
\ea
where $A_{\gen}$ is defined from \nref{variationn} and can be computed using the conical method of \cite{Dong:2013qoa,Camps:2013zua,Fursaev:2013fta}  or directly using the $n\to1$ limit of the Wald entropy (see Section \ref{sec:Wald}). This discussion is independent of how one computes it. To derive the second line, we used that $g_1 +\delta g_{1}$ is a solution to the equations of motion at $n=1$ and we can use the same entropy functional $A_{\gen}$ evaluated on a slightly dislocated surface $X+\d X$. In Section \ref{sec:Wald}, it will be clear how this works when one can take the $\pa_n$ variation inside the action.

Taking the difference of the two equations in \er{variationn} and compare it with \nref{pndi} we get
\begin{equation}
\delta_{\diff} A_{\gen}=0.
\end{equation}
In other words, the entanglement entropy functional should be stationary with respect to shifts in the surface. This argument uses the equations of motion linearized in $n-1$ which is the same condition that led to extremality in \cite{Lewkowycz:2013nqa}.  However, the advantage of our method here is that by considering variations of the action, we do not have to evaluate the equations of motion explicitly.

We expect this to be true for an arbitrary theory of gravity. In the next subsections we discuss the subtleties that lie in these cases.

%%%%%%%%%%%%%%%%%%%%%%%%%%%%%%%%%%%%%%%%%%%%%%%%%%
\subsection{Boundary terms and the $n \rightarrow 1$ limit}\la{secbt}

In the previous discussion, we used the equations of motion at integer $n$ and at the same time deformed the metric off-shell (at finite $n-1$). However, since we want to do two variations of the action, we want to be able to define $\partial_n I(g_n)$ for an slightly off-shell metric, $g_n+\d g_n$.  We want to restrict to ``regular" $\d g_n$: deformations of the metric which give a finite contribution to the action and do not change the strength of the conical singularity. This constraints the variation and allows for a well defined action for the deformed off-shell geometry. 

 We would first like to show that $\delta \hat{I}_n=0$ for all $n$. We can first consider Einstein gravity, where we get
\be\la{einstein}
\delta \hat{I}_n=\lt.\int_{\partial \hat{M}_n} \Theta_{\text{Einstein}}(g_n,\d g_n) \rt|_{r=\epsilon} =\lt.\int_{\partial \hat{M}_n} \sqrt{g_n}  (\nabla^{b} \delta g_{r b}-g_n^{b c}\nabla_{r} \delta g_{b c})\rt|_{r=\epsilon} \overset{\text{regular } \d g_n}{=} 0.
\ee
Because $\sqrt{g_n} \propto r=\epsilon$, it is clear that only if $\d g_n$ diverges approaching the tip one can get a non-zero answer. 

More generally, for an arbitrary higher derivative theory,  we have \cite{Iyer:1994ys}:
\begin{equation}\la{highder}
\delta \hat{I}_n=\lt.\int_{\partial \hat{M}_n} \Theta(g_n,\d g_n) \rt|_{r=\epsilon}=\lt.\int_{\partial \hat M_n} r E_{r b c d} \nabla_c \delta g_{b d} \rt|_{r=\epsilon}
\end{equation}
where $E_{r b c d}$ would be the equations of motion for $R_{a b c d}$, viewed as an independent field. For example for  $f(\text{Riemann})$, $E_{r b c d}=\frac{\partial {\cal L}}{\partial R_{r b c d}}$ . 

It is clear for Einstein gravity that a regular variation of the metric cannot give a finite contribution to the boundary term. However, while \nref{highder} $=0$ at integer $n$, we would also like to argue that this is  true for $1<n<2$. The regularity condition for the variation requires the boundary term \nref{highder} to be finite if not zero. This is because there are no divergent terms at $n=1$ and we are choosing the $\d g_n$ to keep the variation finite for $n >1$. However, the most general metric compatible with replica symmetry will be an expansion with positive powers of $r^{n-1}$ and integer powers of $r$ (see next section). Given that we are working at integer $n$ until the very end,  $\epsilon^{n-1} \rightarrow 0$, which implies that there cannot be a finite term.   This implies that \nref{highder} is zero. 

%%%%%%%%%%%%%%%%%%%%%%%%%%%%%%%%%%%%%%%%%%%%%%%%%%
\subsection{Variational approach for the gravitational entropy} \label{sec:Wald}

While $\partial_n \hat{I}_n|_{n=1}$ in \nref{variationn} can be computed explicitly using squashed cones, that approach requires being careful with several subtleties that arise in the $n \rightarrow 1$ limit and there is currently no complete formula for an arbitrary theory of gravity. In this subsection, we are going to propose an equivalent but perhaps clearer approach than \nref{variationn}, where we think of $\partial_n$ as a variation inside the action. 

Close to the conical singularity, the metric near $n \ap 1$ will schematically look like (we refer the reader to \cite{Dong:2013qoa} for more details):
\ba
g_n &= dr^2+\frac{r^2}{n^2} d\tau^2+(\gamma_n+K_n r^n e^{i \tau}+\cd) dy^2+\cd \eq g_{n;0}+r^{2(n-1)} g_{n;1}+\cd \nonumber\\
\gamma_n &= \gamma_{n;0}+ \gamma_{n;1} r^{2(n-1)}+\gamma_{n;2} r^{4(n-1)}+\cd,\qu
K_n=K_{n;0}+K_{n;1} r^{2(n-1)}+\cd \la{metricansatz}
\ea
where $\cd$ denotes terms which are higher order in $r$, and $\t$ has period $2\pi$. The $n=1$  metric ``splits"\footnote{One way to define $g_{n;0}, g_{n;1}, \cd$ is to require that they contain only integer powers of $r$.}: it is determined by the sum of different terms at $n>1$
\be\la{gspl}
g_{n=1}=g_{1;0}+g_{1;1}+g_{1;2}+\cd.
\ee
 This was seen as a problem for the squashed cone approach in \cite{Miao:2014nxa} (see also \cite{Miao:2015iba,Camps:2016gfs}): in order to determine the ``splitting" one has to solve the most divergent part of the equations of motion, which could be problematic because in order to determine the form of $A_{\gen}$ explicitly one needs the equations of motion at $n \sim 1$.

We would like to understand if we can treat $\partial_n g_n$ outside the $r=\epsilon$ tube as a small variation inside the action integral.  This is not true at $n=1$: the metric might include terms $g_n \propto \epsilon^{2(n-1)}$, which give $\partial_n g_n \propto \epsilon^{2(n-1)}\log \epsilon$, which is not small as $n \rightarrow 1$ (at fixed but small $\epsilon$).  However, we can avoid this issue by working at $n>1$. In this case, we expect that $\partial_n g_n$ is a small variation\footnote{This is true as long as there are no terms in the metric that go like $g \propto\epsilon^{f(n)}$ where $f(n)$ vanishes for any $1<n<n_c$, in which case $\partial_n g$ is a small variation in a finite neighbourhood around $n=1$ (not including $n=1$ itself). We think that this is a very reasonable assumption and we have not been able to find any counterexample. } and thus we can apply \nref{highder} for $\partial_n g_n$. All the contribution from $\partial_n g_n$ comes from the $ g_{\tau \tau}$ component in \nref{metricansatz}. This gives the Wald entropy at finite (but non-integer) $n-1$ :

\begin{equation}
\wtd{S}_n=\partial_n \hat{I}_n=n^{-2} S_{\Wald}(g_{n;0})+\int_{\hat M_n} E_n \partial_n g.
\end{equation}
This formula is valid for non-integer $n$ and it is a finite $n-1$, off-shell version of \nref{variationn}.\footnote{We expect this formula to hold for any $n$ as long as $\partial_n g_n$ is a small variation. If for some reason, the metric splits at some $n_c$, we would define $\wtd{S}_{n_c}=\lim_{n\rightarrow n_c} S_{\Wald}(g_{n;0})$, as we will do in \nref{Agen1}.}  In order to avoid contradictions, it is important that the $n\rightarrow 1$ limit of the Wald entropy at finite $n-1$ is \emph{not} the Wald entropy at the $n=1$ solution. The reason is that the Wald entropy at finite $n-1$ is written in terms of the $g_{n,0}$ fields in \nref{metricansatz}, while at $n=1$ one only have access to the sum over them \er{gspl}. We expect the equations of motion close to $n=1$ to determine $g_{1;0}$ in terms of $g_{n=1}$. 

By carefully taking the limit, one gets the generalized area:
\begin{equation}\la{Agen1}
\wtd{S}_1=A_{\gen}[g_{n=1}]= \lim_{n \rightarrow 1^{+}}S_{\Wald}(g_{n;0})
\end{equation}
where we used the $n=1$ equations of motion. Note that this approach was used before for Einstein gravity in \cite{Lewkowycz:2013nqa,Dong:2016fnf}: because of the simplicity of this theory, one can evaluate \nref{Agen1} directly at $n=1$ without worrying about subtleties in the limit. 

For readers familiar with the squashed cone approach to higher-derivative entanglement entropy \cite{Dong:2013qoa,Camps:2013zua,Fursaev:2013fta}, \nref{Agen1} might look surprising, because $A_{\gen}$ has a contribution from the Wald entropy at $n=1$, but it also has an ``anomalous" contribution which depends on the extrinsic curvature \cite{Dong:2013qoa}. The anomalous contribution depends on the details of how the metric splits.  In our case, $S_{\Wald}$ is explicitly defined in terms of the Lagrangian and the $g_{1;0}$ metric.  In this way, our approach gives an explicit formula for the holographic entanglement entropy: the Wald entropy of the split metric $g_{1;0}$. However, to determine its form in terms of $n=1$ quantities, one has to solve the most divergent part of the equations of motion.

One should be able to show explicitly how \nref{Agen1} relates the squashed cones contribution and the Wald entropy. For Lovelock theories, it is easy to see how this works: $A_{\gen}$ is just given by the Wald entropy in terms of induced Riemann tensor, which is the $n \rightarrow 1$ limit of the projected Riemann tensor on the surface.  For higher derivative theories which have fewer derivatives than Lovelock, such as the one considered in \cite{Castro:2014tta}, we do not have an ``anomalous'' contribution to $A_{\gen}$ and there are no subtleties in taking the $n \rightarrow 1$ limit.  In Appendix \ref{appendixhigher}, we consider a set of two-dimensional examples which we believe capture \nref{Agen1} more generally. 

In our discussion, we have always focused on families of metrics (not necessarily on-shell) which keep the action finite. It is often the case that in the $r \ap 0$ expansion, the most general form for the metric gives rise to an infinite action. In other words, there are some divergent terms in the equations of motion which give a divergent contribution to the gravitational action, while other metric contributions with divergent equations of motion have a finite action (for example, changes in the location of the surface). We will always work with metrics which have a finite action, which is equivalent to imposing the most divergent part of the equations of motion. Even if this class of metrics will depend on the Lagrangian, it is rather universal: it will not depend on the location of the conical singularity. In this way, by requiring the action to be finite, we expect that one can understand the relation between $g_{1;0}$ and $g_{n=1}$, which would determine $S_{\Wald}$ explicitly in terms of $n=1$ quantities.

%%%%%%%%%%%%%%%%%%%%%%%%%%%%%%%%%%%%%%%%%%%%%%%%%%
\section{The first law of entanglement and equations of motion}\label{EOM}

This section is a side product of the previous section. It is independent of the rest of the paper and it will not be mentioned again until the discussion. 
In the previous sections, we have explained how, in classical gravity, the commutativity of the double variation $\partial_n,\delta_{\diff}$ implies the extremality of the entangling functional. We can also use this framework to consider more general variations which do not vanish at the boundary. In holography, it is natural to consider turning on a small source. This framework naturally allow us to derive the integrated equations of motion by assuming that the entanglement entropy is given by the area thus generalizing  \cite{Lashkari:2013koa,Faulkner:2013ica}.

The idea is that, from the field theory perspective, we can think of the second variation commuting as the first law \cite{Blanco:2013joa,Wong:2013gua}: $[\delta,\partial_n] \frac{\log \text{Tr} \rho^n}{n}|_{n=1}=\delta S-\partial_n \text{Tr} \delta \rho \rho^{n-1}=\delta S-\delta K$. We would like to understand if we can recover this from the bulk point of view. 

In order to do this, we want to be in the same setup as \cite{Faulkner:2013ica}. Consider a deformation of the density matrix which changes the one point function of the stress tensor by a small amount, $\delta \langle T_{\mu \nu} \rangle \ll 1$, which is achieved by turning on the respective source,  the boundary metric. If we add a term $\lambda \int d^d x \delta g_{\text{bdy}}^{\mu \nu} T_{\mu \nu}$ to the Lagrangian, then the stress tensor will get an expectation value linear in $\lambda$ (to first order in the deformation). In the original geometry, we expect the same change in the action by computing the variation of the action:
\begin{equation}
\delta_{\lambda} I=\int_M E \delta_{\lambda} g+\lambda\int_{M_{\infty}} d^d x \langle T^{BY}_{\mu \nu} \rangle\delta g_{\text{bdy}}^{\mu \nu}.
\end{equation}

The variation of the action will be given by the equations of motion plus a boundary term, the usual integral of the Brown-York stress tensor. This boundary term will vanish if the expectation value of the stress tensor is zero. 

Now, if we repeat the same for the R\'{e}nyi entropies, we obtain:
\begin{align}
 \delta_{\lambda} \hat{I}_n = & \int_{M_n} E_n \d_{\lambda} g_n
+ \lambda \int_{M_{\infty}} \langle T^{BY}_{\mu \nu}\rangle_n \delta g_{\text{bdy}}^{\mu \nu}.
\end{align}

We can analytically continue this expression in $n$, take its $n$ derivative, and express it in terms of boundary quantities using the standard dictionary $\langle T^{BY} \rangle=\langle T \rangle$:
\begin{equation}
\partial_{n} \delta_{\lambda} \hat{I}_n|_{n=1}=\lambda\int d^d x \langle K T_{\mu \nu}\rangle\delta g_{\text{bdy}}^{\mu \nu}+\partial_n \int_{M_n} E_n \d_{\lambda} g_n|_{n=1}=\delta_{\lambda} \langle K \rangle+\partial_n \int_{M_n} E_n \d_{\lambda} g_n|_{n=1}.
\end{equation}

This formula for the variation of the boundary Hamiltonian from analytically continuing the one point function at integer $n$ was discussed previously in \cite{Rosenhaus:2014woa,Lewkowycz:2014jia}. Note that in the case where the modular Hamiltonian is local, the right-hand side (RHS) will be given by $\int_{R} d\Sigma^{\mu} \xi^{\nu} \delta\langle T_{\mu \nu} \rangle$ and this can be understood from the left-hand side (LHS) because $\delta \langle T_{\mu \nu} \rangle=\int d^d x \langle T_{\mu \nu} T_{\alpha \beta}\rangle\delta g_{\text{bdy}}^{\alpha \beta}$.  So we are in exactly the same setup as \cite{Faulkner:2013ica}. 

We can try to understand the variations in the opposite order:
\begin{equation}
\partial_n \hat{I}_n|_{n=1}=\int E \partial_n g +  A_{\gen} \rightarrow \delta_{\lambda} \partial_n \hat{I}_n|_{n=1}=\delta_{\lambda} \int E \partial_n g +\delta_{\lambda} A_{\gen}
\end{equation}
where we have not yet used any equation of motion.

In this way, given that the variations commute with each other, we obtain:
\begin{equation}
[\partial_n,\delta_{\lambda}] \hat{I}_n|_{n=1}=\delta_{\lambda} \langle K \rangle-\delta_{\lambda}  A_{\gen}-\int \delta_{\lambda} E \partial_n g|_{n=1}+\int \partial_n E_n|_{n=1} \delta_{\lambda} g.
\end{equation}
We have derived this equation by assuming that there is some action, but this equation should be a true equation independently of how we derive it. Note that, to derive it, we did not need to use the background equations of motion since they cancel in the double variation.

This gives a gravitational entanglement first law, in a very similar to Wald's first law \cite{Iyer:1995kg}. In both cases one derives the first law by varying the Lagrangian. In Wald's case, the first law is a consequence of having a Killing vector: the conservation of diffeomorphism current relates the difference between the area in the extremal surface and the energy at infinity with the gravitational constraints, integrated in a Cauchy slice in the entanglement wedge. In our case, we do a $\partial_n$ variation, which is less symmetric and we obtain that the two boundary terms differ by a codimension $0$ integral. In this way, under the assumptions that the entanglement entropy is given by the generalized area and that the  background equations of motion are satisfied close to $n=1$, we have derived the following equation:
\begin{equation}
\delta_{\lambda} S-\delta_{\lambda} K =\int \delta_{\lambda} E \tilde{g}
\end{equation} 
with $\tilde{g}=\partial_n g$,  but the equation is true even if we do not know what $\tilde{g}$ is.  In this case, $\delta E$ is integrated over the whole manifold. Since we have less symmetries that in Rindler (where there is a Killing vector), the integral is higher dimensional, but it does not seem possible to do better from the first law. 

From the assumptions that the background metric to satisfy the background equations of motion at leading order in $n-1$, the standard bulk-boundary dictionary and that the entropy is given by the area, we have deduced that $\delta S=\delta K \iff \int \delta E \tilde{g}=0$. Since this is true for an arbitrary entangling surface, this probably implies $\delta E=0$ everywhere. In principle, the linearized equations around an arbitrary background could be integrated to give the nonlinear equations of motion. However, given that the leading order in $(n-1)$ background equation of motion is a necessary assumption for this discussion, one might need to assume the background equations of motion for all $n$ to derive the nonlinear Einstein equations.\footnote{This is because the equivalent of the first law for the modular entropies would give us the linearized equation of motion at arbitrary $n$ from which the nonlinear equation can be obtained.}

Note also that this expression for the modular Hamiltonian is compatible with \cite{Jafferis:2015del}. In fact, for Einstein gravity, we can think of $\delta A=\int_{RT} \gamma^{\alpha \beta}\delta g_{\alpha \beta}$ and express $\delta g=\int_{M_{\infty}} dx G(X,x) T(x)$. This gives an expression for $\delta K$ from which we can read $\langle K T_{\mu \nu} \rangle$ in holographic theories (similar comments were made in \cite{Blanco:2013joa,Jafferis:2014lza}). The reason why this is only true given the equations of motion is because in order to write the metric operator in terms of the boundary fields one imposes the linearized equations of motion for the graviton. The good thing about the euclidean prescription described above is that it provides a bulk definition for the modular Hamiltonian which is independent of the area  through the asymptotic one point functions at $n \sim 1$. 

%%%%%%%%%%%%%%%%%%%%%%%%%%%%%%%%%%%%%%%%%%%%%%%%%%
\section{Quantum corrections to entanglement entropy}\label{qext}
In the presence of quantum corrections, we will have a path integral in the replicated space $M_n$. The presence of quantum corrections will modify the equations of motion to all orders in $G_N$, we are going to denote the backreacted background metric by $g_{cl,n}$ and will expand it in $G_N$: $g_{cl,n}=g_{cl,n}^{(0)}+G_N g_{cl,n}^{(1)}+\cd$.\footnote{By the label classical, we mean that it is not a fluctuating but rather a background field. $g_{cl,n}$ will contain $G_N$ corrections due to the backreaction of the quantum fields.} As in \cite{Lewkowycz:2013nqa}, we assume that the background metric $g_{cl,n}$ is $\bZ_n$ symmetric. 

We are going to define the ``orbifolded" partition function by dividing by $n$:
\begin{eqnarray}
-\log Z_n=I_{grav}(g_{cl,n})-\log Z^{matter}_n(g_{cl,n}), ~~ \hat{I}_{n}[g_{cl,n}]=-\frac{1}{n} \log Z.
\end{eqnarray}

Let us review the discussion of \cite{Faulkner:2013ana}, where they describe how to think about $\log Z_n$, $\partial_n \hat{I}_n$ at non-integer $n$. In the previous classical discussion, because of the $\bZ_n$ symmetry of the background, the calculation of the action only needed the metric in the quotient space, however the quantum partition function is only defined  in the parent space.\footnote{This is just the statement that the background metric is a one point function, which is $\bZ_n$ invariant, while higher correlators need the whole parent space.} We can exploit the $\bZ_n$ symmetry of the background metric, to write the partition function as:
\begin{equation}\label{Zq}
\log Z_n=\log \text{Tr} \rho_n^n
\end{equation}
where the gravitational density matrix $\rho_n$ is defined by the boundary condition that the background metric $g_{cl,n}$ has a conical singularity of strength $1/n$. By taking $n$ powers of this seemingly singular density matrix, one ends up with a geometry which does not have a conical singularity. Given that $\rho_n$ is defined for arbitrary $n$, one can analytically continue \nref{Zq} to real $n$: it is just the $n$-th power of $\rho_n$. In this way, we can express the derivative of $\hat{I}_n$ as the sum of the derivatives with respect to the lower and upper arguments of $\text{Tr} \rho_n^n$:
\begin{eqnarray}\label{nderiv0}
\partial_n \hat{I}_n =-\partial_{\delta n} \log \text{Tr} \rho_{n+{\delta n}}^n -\partial_{\delta n} \frac{1}{n+\delta n} \log \text{Tr} \rho^{n+{\delta n}}_n=n^{-2}\langle S_{\Wald} \rangle_n+\tilde{S}_{n,\bulk}.
\end{eqnarray}

These first term is obtained by taking a derivative with respect to the background metric inside the path integral and using the expectation value of the equations of motion as in \cite{Faulkner:2013ana} (but to all orders in $G_N$).  To exploit the semiclassical part of the problem (which allowed us to use the $\rho_n$ notation), where we have a well defined background metric, one needs to work perturbatively in $G_N$ around a given saddle $g_{cl,n}^{(0)}$. This discussion only makes sense in the $G_N$ expansion. This formula is formally true for arbitrary $n$, however to get the corrections to the background metric $g^{(k>0)}_{cl,n}$ one needs to analytically continue the expectation value of the stress tensor, $\langle T \rangle_n$, to non-integer $n$. 

We can take the $n \rightarrow 1$ limit:
\begin{eqnarray}
S=\lim_{n \rightarrow 1^{+}} \left ( n^{-2}\langle S_{\Wald} \rangle_n+\tilde{S}_{n,\bulk} \right )= \langle A_{\gen} \rangle+ S_{\bulk}=S_{\gen} \label{eq:qext}.
\end{eqnarray}
To one loop, this is the same as \cite{Faulkner:2013ana}. The notation is a little different. There, $\langle A_{\gen} \rangle$ was explicitly separated into two terms: one coming from the generalized area evaluated in the background metric $g_{cl}$ (which was denoted $\frac{\delta A}{4 G_N}$) and a contribution coming from matter fields which couple with derivatives of the metric,  $\langle S_{wald-like} \rangle$. This last term is easily illustrated with a scalar field with a term $\int R \phi^2$, where $S_{wald-like}=\int_{RT} \phi^2$. In this original notation, the expectation value of the area due to graviton fluctuations should be thought as included in $S_{wald-like}$.

This procedure is in principle well defined to all orders in $G_N$:  $\log Z_n$ is a completely standard partition function, although equation \nref{nderiv0} requires introducing a $r=\epsilon$ artificial boundary in our gravitational background. This ``brick wall" partition function has been discussed in detail in 
\cite{Donnelly:2014fua,Donnelly:2015hxa}.

More concretely, at integer $n$, the partition function is well defined and nothing special happens at the $\mathbb{Z}_n$ symmetric fixed point. In order to take the $n$ derivative, it is convenient to define the partition function with a boundary at $r=\epsilon$. We want to do this in a way that we recover the original
partition function when $\epsilon \rightarrow 0$. This is achieved by choosing a set of boundary conditions for the
quantum fields at $r=\epsilon$ and then integrating independently over
all possible boundary conditions. This integration is often referred
to as summing over edge modes 
\cite{Donnelly:2014fua,Donnelly:2015hxa}, there they
write the partition in a smooth black hole background for abelian gauge fields
in terms of the partition function in a brick wall geometry summed
over all possible electric fluxes across the boundary.  Of course,
after setting up these boundary conditions to define the partition
function in the presence of a boundary, the entropy ($n$ derivative)
will also have the same boundary conditions and edge modes. We can
think of these edge modes as the center variables of
\cite{Casini:2013rba}. We expect this story to generalize
straightforwardly to gravity, see \cite{Jafferis:2015del} for a discussion about gauge
invariant boundary conditions for free gravitons.

%%%%%%%%%%%%%%%%%%%%%%%%%%%%%%%%%%%%%%%%%%%%%%%%%%
\subsection{Variations}

In order to take variations with respect to the background metric, we have to define our partition function slightly off-shell. 
We can do this by adding a background stress tensor which couples with the metric operator: $\int dx^d\sqrt{g} T_{\mu \nu}^{\text{bkg}} h^{\mu \nu}$, with $h^{\mu \nu}=g^{\mu \nu}-g_{cl,n}^{\mu \nu}$, the background subtracted metric, it is hopefully clear from context that $h,g$ denote operators while $g_{cl}$ is a c-number. This term in the Lagrangian naturally splits the metric operators into the background metric, $g_{cl,n}$ and background subtracted fluctuation, which we will denote by $h$.\footnote{To each order in $G_N$, we can think of the Einstein equation as simply the tadpole equation for the metric operator: $g_{cl,n}=\langle g \rangle_n$.} Derivatives with respect to the background stress tensor generate then background subtracted metric correlations. The role of the background stress tensor is to turn on-shell an arbitrary background metric\footnote{We can think of $T^{\text{bkg}}=T^{\text{bkg}}[g_{cl}]$, since the equations of motion (tadpole equations) are $ E(g_{cl})-G_N \langle T \rangle=  T^{\text{bkg}}$ and the LHS defines $T^{\text{bkg}}[g_{cl}]$. Equivalently, we can do Legendre transformation and obtain the effective action, which is a function of the off-shell background metric. } which allows us to think of the partition function as  a function of the background metric.

At integer $n$, we will consider the variation of $\hat{I}$ with respect to the background metric:
\begin{eqnarray}\delta \hat{I}_{n}|_{T_{\text{bkg}}=0}=\int_{\hat{M}_n} (E_n(g_{cl,n})+\langle T \rangle_n) \delta g +\int_{\partial \hat{M}_n} \Theta(g_{cl,n},\delta g)|_{r=\epsilon} =0
\end{eqnarray}
where we used the quantum corrected equations of motion and the results from the previous section.  Since this equation is valid for arbitrary $n$, its $n$ derivative will be zero. The boundary term appears when $g_{cl,n}$ has to be integrated by parts and it should be thought as including an expectation value with respect to the fluctuating fields, but we omitted it to simplify the notation.

By turning a background stress tensor, we can also take variation of \nref{nderiv0}
\begin{eqnarray}\label{nderiv}
\delta \partial_n \hat{I}_n|_{T^{\text{bkg}}=0} =n^{-2} \delta \langle  S_{\Wald} \rangle_n+\delta \tilde{S}_{n,\bulk} +\int_{\hat{M}_n}(\delta E(g_{cl,n})+\delta \langle T \rangle_n) \partial_n g.
\end{eqnarray}

As our variation would be off-shell at integer $n$, the last term will not cancel. However, if we consider a variation which is on-shell close at $n=1$, a diffeomorphism, the variation of last term will be zero, so, asking for $\delta \partial_n \hat{I}_n=\partial_n \delta \hat{I}_n=0$ implies that
\begin{equation}
\delta_{\diff}(\langle A_{\gen} \rangle+S_{\bulk})=\delta_{\diff} S_{\gen}=0.
\end{equation}

This is the quantum extremality condition of \cite{Engelhardt:2014gca}. 
To leading order in $G_N$, we will later show explicitly that this is true using the equations of motion at $n \sim 1$, but this approach is valid to higher orders in $G_N$. An example with finite backreaction would be that of the Polyakov action (see Appendix \ref{appendixPoly}), but this example might be too simple, since its effective action is local. 

%%%%%%%%%%%%%%%%%%%%%%%%%%%%%%%%%%%%%%%%%%%%%%%%%%
\subsubsection*{$G_N$ perturbation theory, the stress tensor and gravitons}
The previous discussion applied order by order in $G_N$ and here we will be more a little bit more explicit about how it is defined. 

The Einstein equation is an operator equation, which means that:
\begin{equation}
\langle E(g) \rangle=G_N \langle T_{matter}(g,\phi) \rangle.
\end{equation}

 We can expand the Einstein tensor in terms of $g=g_{cl}+h$ in $G_N$ and, to each order, we can basically think of the gravitons $h$ as interacting matter with an their effective stress tensor determined by the expectation value of the Einstein tensor, expanded around with $E(\langle g \rangle)$. In this way,  we can write the $O(G_N^k)$ term in the previous equation as:
 \begin{equation}
 E^{lin}_n(g^{(k)}_{cl})+E(g^{(j<k)}_{cl})|_{O(G_N^k)}=\left[ G_N \langle T_{matter}(g_{cl},h,\phi) \rangle+\langle T_{grav}(h,g_{cl}) \rangle \right]_{O(G_N^k)}=G_N \langle T \rangle|_{O(G_N^k)} \label{Eingrav}
 \end{equation}
where the first term in the LHS is the linearized Einstein tensor  and this equation determines $g_{cl,n}^{(k)}$ in terms of expectation values and $g_{cl,n}^{(j<k)}$  and can be thought as a tadpole contribution to $g_{cl,n}^{(k)}$.  Note that $T_{grav}$ is defined order by order in $G_N$ by expanding $\langle E_n(g) \rangle$. We schematically denote the RHS as $\langle T \rangle$. 

We can think of the equations of motion as a background field expansion of the action order by order in $G_N$ and consider the variation of the (effective) action with respect to the background metric. If we think about gravitons order by order, they are basically the same as complicated matter with an effective stress tensor determine by the previous equation. 

%%%%%%%%%%%%%%%%%%%%%%%%%%%%%%%%%%%%%%%%%%%%%%%%%%
\subsection{The definition of quantum extremal surfaces} \label{definitionqext}

In the previous sections, we derived the quantum extremality condition. In this section, we will explore the quantum extremality equations. Note that, in order to have a non-trivial quantum extremal surface, there has to be some asymmetry between the inside and outside region, and, for the symmetric case of a sphere in the vacuum, there will not be corrections to the extremal surface. 

In our framework, we will always have a well defined background metric $g_{cl}$ and interacting gravitons on top of it.  We can think of the location of the entangling surface in similar terms: $X=X_0+G_N X_1+\cd$, $X$ denotes the location of the surface to all orders.\footnote{Note that there are no $G_N^{1/2}$ contributions since the entanglement entropy from gravitons is $O(G_N^0)$} For Einstein gravity (it generalizes trivially to higher derivatives but we are going to focus on Einstein for simplicity), the leading term corresponds to the location of the extremal surface \begin{equation}
\frac{1}{4 G_N} {\cal K}_{X_0}^I(g^{(0)}_{cl};y)=0+O(G_N^0),
\end{equation}
where ${\cal K}$ is the extrinsic curvature of the surface at $X_0$ and it depends on the position on the RT surface $y$ and in the background metric, since it is codimension $2$ surface, there are two normal directions which we denote by $I$.  To leading order in $G_N$, we can write an equation for the quantum extremal surface using the results of \cite{Rosenhaus:2014woa,Allais:2014ata}. One can use perturbation theory to understand how the entropy changes by a small change in the subregion. As in the previous discussion, we are going to denote by $r=\epsilon$ the tubular region close to the entangling surface. Using their work, one can show that to first order in $G_N$:
 \begin{eqnarray}
  \frac{1}{4 G_N}{\cal K}_{X_0+X_1}^I(g_{cl};y)\big{|}_{O(G^0_N)}=-\delta_{X^I} S_{\bulk}(X_0) =-2 \pi \lim_{\epsilon \rightarrow 0}\epsilon \langle T^{I r}(r=\epsilon;y) K_0 \rangle.
\end{eqnarray}
This is a linear equation for $X_1$, determined in terms of quantities evaluated at $X_0$ (the classical extremal surface) which are well defined.  $T$ is the RHS of \nref{Eingrav} and it is evaluated $\epsilon$ away from the entangling surface.  The finite contribution to the variation of the entanglement entropy comes from a divergent contribution of $\langle T K \rangle$. In general terms, we expect this object to diverge when the stress tensor approaches the boundary of the region and the leading divergence goes like $\frac{1}{{\epsilon}^{d-2}}$. All the contributions that give a divergent variation of the entropy will correspond to the renormalization of the gravitational couplings, and should disappear after adding the proper counterterms. So, only the divergent contribution  $\langle T K \rangle \propto \frac{1}{\epsilon}$ will contribute. If the background has a Killing vector, this correlator will not have an odd divergent term. The higher orders can be obtained from solving the exact equation
\begin{equation}\label{qextx}
\frac{1}{4 G_N} {\cal K}_X^I(g_{cl};y)=- 2 \pi \lim_{\epsilon \rightarrow 0}\epsilon \langle T^{I r}(r=\epsilon;y) K^X\rangle 
\end{equation}
 where $K^X$ is the modular Hamiltonian of the bulk surface $X$. This equation can be expanded in $X$ order by order in $G_N$. Of course, ${\cal K}$ should be thought as an expectation value and \nref{qextx} as a tadpole equation for $X$, for example to $O(G_N^0)$, we can think of adding an extra term in the RHS $-\langle {\cal K}_{X_0}(h;y) \rangle$. 

To leading order in $G_N$, we can also see how one would obtain the quantum extremality condition from the equations of motion around $n \sim 1$. The extremality of the area in RT is obtained by expanding the equations of motion near $n \sim 1,r\sim 0$ \cite{Lewkowycz:2013nqa}. Schematically:
\begin{equation}
E_n=0 \times (n-1)^0+(n-1) \frac{{\cal K}^I(y)}{r}+O( (n-1)r^0) +O((n-1)^2) \rightarrow {\cal K}^I(y)=0,
\end{equation}
that is, extremality is derived from regularity of the metric close to the $\bZ_n$ symmetric fixed point. In the presence of quantum matter, we will have:
\begin{equation}
E_n-G_N\langle T \rangle_n=0 \times (n-1)^0+(n-1) (\frac{{\cal K}(y)}{r}+ 8 \pi G_N \langle T K \rangle)+\cd.
\end{equation}
It is now clear that if there is a $1/r$ divergent term in $\langle T K \rangle$, regularity of the metric close to the $\bZ_n$ symmetric fixed point will shift the surface to the quantum extremal. It is also clear from this equation that the stress tensor that appears in \nref{qextx} is just the RHS of Einstein equations. 

%%%%%%%%%%%%%%%%%%%%%%%%%%%%%%%%%%%%%%%%%%%%%%%%%%
\subsubsection*{Subtleties with gravitons} 

It might not be completely clear how to evaluate the entanglement entropy in the quantum extremal surface for gravitons, or whether it is well defined (see \cite{Jafferis:2015del} for a set of boundary conditions that works for extremal surfaces). We certainly expect $\log Z_n$ to be well defined to all orders in $G_N$ and $g_{cl,n}$ should also be well defined in the $G_N$ expansion. Upon the inclusion of a boundary and summing over the proper edge modes, we expect that  (\ref{eq:qext},\ref{qextx}) makes sense order by order in $G_N$. 

Of course, in order to make this more concrete one should understand better the entanglement entropy of gravitons. For free gravitons, we expect that one can apply the ideas of \cite{Donnelly:2014fua,Donnelly:2015hxa} together with  \cite{Jafferis:2015del} to compute the entanglement entropy. Then,  we expect that the interacting graviton can be treated in the same way, by considering the interaction in entanglement perturbation theory \cite{Rosenhaus:2014woa,Faulkner:2015csl,Faulkner:2016mzt}. In the same way, we expect that the deformation of the surface away from extremality can be understood in similar terms.  More explicitly, as long as the displacement is small, we will schematically have
\bm
S_{\bulk}=\sum_m \int_{RT} dy_1\int ds_1  \cd \int_{RT} dy_m  \int ds_m \delta X(y_1) \cd \delta X(y_m)\times\\
\times \langle T_s(X_0,y_1) \cd T_{s_m}(X_0,y_m) f(K_{0})   \rangle, 
\em
with $T_s=e^{i K_0 s} T e^{-i K_0 s}$, the modular evolved stress tensor. That is, the bulk entanglement entropy in a neighboring surface  will be a correlator of (modular evolved) stress tensors and some function of the modular Hamiltonian $K_0$ integrated several time over the extremal surface. So, in principle, we might only need the modular Hamiltonian in extremal surface to obtain the entanglement entropy in other surfaces. In this expression, part of the $G_N$ will come from $\delta X$, part from changing the background metric and part from the correlator: stress tensors and $K_0$, for example to $O(G_N)$ we will have $S_{\bulk}=S_{\bulk,free}(X_0)+S_{\bulk,G_N}(X_0)+\int_{RT} dy \delta X^I \langle T^{I r}(X_0,y) K_0 \rangle +  \int dx \delta g_{a b} \langle T^{a b}(x) K_0 \rangle$.

Alternatively, we could just define this graviton entanglement entropy in terms of the boundary replica trick. We expect the partition function in this smooth manifold to be perfectly well defined. 

Note that quantum extremality relates the contributions from $\delta X$ of $S_{\bulk}$ to the contribution from the area. We will discuss this more explicitly in the next section.

%%%%%%%%%%%%%%%%%%%%%%%%%%%%%%%%%%%%%%%%%%%%%%%%%%
 \subsection{Quantum extremality and mixtures}
 
 Up to here, we have discussed quantum extremality in terms of partition functions $\text{Tr} \rho^n$ which have a well defined path integral preparation and correspond to a unique classical saddle. We would like to understand how to extend the previous methods to mixtures of states:
 \begin{equation} \label{supZ}
Z_{n}[\rho+\sigma]= \text{Tr} (\rho+\sigma)^n=\sum_{A_k=\lbrace \sigma,\rho \rbrace}  \text{Tr} (A_1 A_2 A_3 \cd A_n).
 \end{equation}
 
 Even if $Z_{\rho+\sigma,n}$ cannot be prepared in the Euclidean path integral, each of the terms in the RHS of \nref{supZ} can, so we can think of \nref{supZ} as a sum of path integrals. So, $Z_{\rho+\sigma,n}$ is in principle well defined for integer $n$: we have an asymptotic circle with perimeter $2\pi n$ which is divided into $n$ slices and set boundary conditions in each of the slice determined by a configuration in the RHS of \nref{supZ}. Because this definition is an $n-$dependent sum of path integrals, it seems hard to analytically continue in $n$.

At this point, it is useful to make a remark about mixtures of path integrals in general. In the effective action formalism that we described before, whenever we have a mixture of states, we want to fix the same background source across the different states.  Since there is only one background source, there is only one corresponding classical value for the field, that is, one tadpole equation.   Consider the example of the linear mixture of two density matrices: $\frac{1}{2}(\rho+\sigma)$. In the presence of the same background tensor $\frac{\delta}{\delta T^{\text{bkg}}} (Z_{\rho}+Z_{\sigma})=\langle g \rangle_{\rho}+\langle g \rangle_{\sigma}=2 g_{cl,\rho+\sigma}$, we can Legendre transform by adding the term $-\int dx^d T^{\text{bkg}} g_{cl,\rho+\sigma}$ to the path integrals. This means that $\frac{\delta}{\delta g_{cl,\rho+\sigma}} Z=0$ will give the sum of the equations of motion, the tadpole condition will be $\langle  g-g_{cl,\rho+\sigma} \rangle_{\rho+\sigma}=0$ which is not explicitly linear in $\rho+\sigma$ because it is expanded around a background. If $Z_{\rho}$ and $Z_{\sigma}$ share the same saddle to leading order in the saddle point expansion\footnote{If $g_{cl,\rho}^{(0)} \not = g_{cl,\sigma}^{(0)}$, $g_{cl,\rho+\sigma}$ is not a saddle. While $g_{\rho+\sigma}$ appears when coupling of the two path integrals through a background stress tensor, it does not have a clear semiclassical interpretation and we will not be considering this situation.}, $g^{(0)}_{cl,\rho}=g^{(0)}_{cl,\sigma}$, then we can understand this formalism as adding a quantum mixture of states to the classical geometry and solving the sum of the equations of motion $E(g_{cl,\rho+\sigma})=\langle T^{g_{cl,\rho+\sigma}}\rangle_{\rho+\sigma}$, which we can now compute in $G_N$ perturbation theory.\footnote{This gives a well-defined procedure to compute the partition functions. If the two states are macroscopically distinguishable, the gravitons $h=g-g_{cl,\rho+\sigma}$ would not have a well-controlled one-loop partition function. However, for $h=g-g_{cl}^{(0)}-G_N g^{(k>0)}_{cl,\rho+\sigma}$, this graviton is only slightly off-shell with respect to the path integral of $\rho$ or $\sigma$, so the difference is small and it has a well-defined partition function. Alternatively, we can compute these partition functions with respect to their on-shell background first and use linearity $g_{cl,\rho+\sigma}=g_{cl,\rho}+g_{cl,\sigma}$.}  Note that even if the Einstein equations $\langle E(g) \rangle=\langle T_{matter} \rangle$ are linear in the mixture, the expectation value of the tadpole is background dependent. This makes the linearity of Einstein equations hard to see if we write them around $g_{cl,\rho+\sigma}$, however it is clear that $g_{cl,\rho+\sigma} = \frac{1}{2}(g_{cl,\rho}+g_{cl,\sigma})$  (yet this is clear because $\langle g \rangle_{\psi}=g_{cl,\psi})$.   

The previous discussion gives a prescription to extend our result to mixtures of states that have the same $O(G_N^0)$ value of the metric: $g^{(0)}_{cl,\rho}=g^{(0)}_{cl,\sigma}$. These two states share, to leading order in $G_N$, the same ($\bZ_n$ symmetric) saddle for $Z_n$. We can think of the sum of path integrals in terms of a mixture of quantum states in the $g_{cl}^{(0)}$ geometry, satisfying the equations of motion:
\begin{equation}\label{superEOM}
E(g_{cl,(\rho+\sigma)^n})=\langle T^{g_{cl,(\rho+\sigma)^n}}\rangle_{(\rho+\sigma)^n}
\end{equation}
 where we think of the RHS as a sum over partition functions and the superscript denote that we are expanding the gravitons around the $g_{cl,(\rho+\sigma)^n}$ background.  It is key that we phrase the problem in terms of a unique geometry and not a mixture of them since this will allows us to analytically continue in $n$. To do this, we note that $g_{cl,(\rho+\sigma)^n}$ is $\bZ_n$ symmetric, which allows us to think of $\text{Tr} (\rho+\sigma)^n$ in terms of taking the $n$-th power of $(\rho+\sigma)_n$, where the subscript $n$ denotes that it has the metric determined by \nref{superEOM}.  Upon analytic continuation of the RHS of \nref{superEOM}, the previous gives a prescription to compute $Z_n(\rho+\sigma)=\text{Tr} [(\rho+\sigma)_n]^n$ for non integer $n$. Given this, the discussion from the previous section follows and we get quantum extremality for mixtures:
 \begin{equation}
 \frac{1}{4 G_N}{\cal K}_X^I(g_{cl,\rho+\sigma};y)=-2 \pi \lim_{\epsilon \rightarrow 0} \epsilon \langle T^{I r} (r=\epsilon,y) K^X_{\rho+\sigma} \rangle_{\rho+\sigma}.
 \end{equation}
It is clear that we want to think of the $n=1$ solution as given by a unique geometry, $g_{cl,\rho+\sigma}$ where quantum states can be entangled. Note that the fact that at integer $n$ we have complicated sums of partition functions makes the quantum extremal surface nonlinear in the state, since it depends on the modular Hamiltonian of the mixture, ie $X_{\rho+\sigma} \not = X_{\rho}+X_{\sigma}$ because $g_{cl,(\rho+\sigma)^n} \not = g_{cl,\rho^n}+g_{cl,\sigma^n}$.
 
%%%%%%%%%%%%%%%%%%%%%%%%%%%%%%%%%%%%%%%%%%%%%%%%%% 
\section{Modular extremality}\label{mext}
 
 A simple consequence of quantum extremality for mixtures is that we can compute the expectation value of modular Hamiltonians for states close to each other (same $g_{cl}^{(0)}$).  The modular Hamiltonian is just the log of the density matrix :
\begin{equation}
\langle K_{\sigma} \rangle_{\rho}=-\langle \log \sigma \rangle_{\rho}=-\partial_n \langle \sigma^{n-1} \rho \rangle.
\end{equation} 
 
 We can get this from a mixture $\sigma+\lambda \rho$, since $\partial_{\lambda} \text{Tr} (\sigma+\lambda \rho)^n|_{\lambda=0}=\text{Tr} \rho \sigma^{n-1}$. 

In this way, if we combine this with quantum extremality for mixtures\footnote{For small perturbations, one can actually derive modular extremality for $\langle K_{\sigma} \rangle_{\sigma+\delta \sigma}$ in terms of quantum extremality plus the first law for $S(\sigma+\delta \sigma)=\langle K_{\sigma+\delta \sigma} \rangle_{\sigma+\delta \sigma}$.}, we get a formula for the dual to the modular Hamiltonian:
\begin{equation}
\langle K_{\sigma} \rangle_{\rho}=\frac{\langle A^{X_{\sigma}} \rangle_{\rho}}{4 G_N}+\langle K_{\bulk,\sigma}^{X_{\sigma}} \rangle_{\rho}, ~~~ \frac{\delta}{\delta {X_{\sigma}}} \langle K_{\sigma} \rangle_{\rho}=0.
\end{equation}

The boundary modular Hamiltonian is just given by the area plus the  expectation value of the bulk modular Hamiltonian of $\sigma$ in the $\rho$ background. For simplicity of notation, we will illustrate this for Einstein gravity, but it generalizes trivially to higher derivatives. The surface where these terms are evaluated is determined by quantum extremality for the mixture, which implies that the sum of the two terms is extremized. We will call the $X_{\sigma}$ surface \emph{modular extremal}. The variation can be carried out \cite{Faulkner:2016mzt}:\footnote{Note the first term in this expression is only discussed in their appendix since in they focus on the full modular Hamiltonian, where such a term is not present.}
\bm
\frac{1}{4 G_N}{\cal K}_{X_{\sigma}(\rho)}^I(g_{cl},y) =-\frac{\delta}{\delta X_{\sigma}^I} \langle K^{X_{\sigma}}_{\bulk,\sigma} \rangle_{\rho} \\
=\lim_{\epsilon \rightarrow 0} \epsilon \left[ \langle :T^{I r}(r=\epsilon;y):_{\rho} K^{X_{\sigma}}_{\bulk,\sigma} \rangle_{\rho} +   \int_{-\infty}^{\infty}  \frac{ds}{4 \sinh^2 (s/2+i \varepsilon)} \langle :T_s^{I r}(r=\epsilon;y):_{\sigma}  \rangle_{\rho} \right] \label{modext}
\em
where $:T:_{\rho}=T-\langle T \rangle_{\rho}$ and  $:T_s:_{\sigma} \equiv \exp(i K_{bulk,\sigma}^{X_{\sigma}} s) : T : _{\sigma} \exp(-i K_{\bulk,\sigma}^{X_{\sigma}} s)$. As we discussed before, one should also add an expectation value of the extrinsic curvature for the gravitons in the RHS but we omitted it for simplicity. The finite contribution arises from a $1/\epsilon$ divergence in the first term, as in quantum extremality, and the second term can in principle get finite contributions from the $s$ integral (for a local modular Hamiltonian  there are contributions from $s \sim -\log \epsilon$ that make this finite). We can think of the first term of the variation of $\rho$ and the second term the variation of $\log \sigma$. If $\rho=\sigma$, then the second term does not contribute, since it is proportional to the one point function of the stress tensor and we recover quantum extremality for a single state.

To leading order in $G_N$, $X_{\sigma}(\rho)$ is just the classical extremal surface and this is the bulk expression for the modular Hamiltonian discussed in \cite{Jafferis:2015del}. In that paper, it was also discussed what the dual of the relative entropy is to leading order in $G_N$ and our result generalizes it to higher orders:

 \begin{equation}
 S_{\rel}(\rho|\sigma) \equiv \langle K_{\sigma} \rangle_{\rho}-\langle K_{\rho} \rangle_{\rho}= \langle \frac{A^{X_{\sigma}}}{4 G_N}-\frac{A^{X_{\rho}}}{4 G_N}+K^{X_{\sigma}}_{\bulk,\sigma}-K_{\bulk,\rho}^{X_{\rho}} \rangle_{\rho}.
 \end{equation}

Given that the surfaces where the modular Hamiltonians are evaluated are different, the relative entropy does not have a simple description. Its difference from the bulk relative entropy can be understood as coming from the difference in areas localized $O(G_N)$ away from the classical extremal surface. From our point of view, the object which has a natural bulk description is the modular Hamiltonians, since it has a  well defined path integral.

%%%%%%%%%%%%%%%%%%%%%%%%%%%%%%%%%%%%%%%%%%%%%%%%%%
\subsection{A linear mapping of surfaces}

From the point of view of the path integral at integer $n$, $\text{Tr} \rho \sigma^{n-1}$, it is clear that our expression should be linear in $\rho$. At $n=1$, this is the statement that we should be thinking of the position of the modular extremal surface $X_{\sigma}(\rho)$ as linear function of the state $\rho$. 

In this way, given a state $\sigma$ and its quantum extremal surface $X_{\sigma}(\sigma)$ , we can think of $X_{\sigma}(\rho)$ as a mapping from the quantum extremal surface in the $\sigma$ background to a surface in the $\rho$ background  (this is similar to \cite{Dong:2016eik}, where some unspecified mapping was proposed).

The $G_N$ corrections generalize the extremal area operator appearing in \cite{Jafferis:2015del} to the $\sigma$-dependent modular area operator: $A^{X_{\sigma}}$ depends on the modular Hamiltonian of $\sigma$. Since our equation can be understood as the expectation value of an operator in the state $\rho$, we can write is as an operator equation:

\begin{equation}
K_{\sigma}=\frac{A^{X_{\sigma}}}{4 G_N}+K^{X_{\sigma}}_{\bulk,\sigma}.
\end{equation}

\subsection*{Linearity and state dependent divergences}

In principle, one could worry about the fact that \nref{modext} is not linear in $\rho$ because of
\begin{equation} \label{1stterm}
\lim_{\epsilon \rightarrow 0} \epsilon \left[ \langle T^{I r}(r=\epsilon,y) K_{\bulk,\sigma} \rangle_{\rho}-\langle T^{I r}(r=\epsilon,y)  \rangle_{\rho} \langle K_{\bulk,\sigma} \rangle_{\rho} \right ].
\end{equation} 
Note however that in the second term, the divergent contribution has to come from $K_{\bulk,\sigma}$, since there is nothing special happening at $r=\epsilon$ in the original state. Now, if this divergent contribution from $K_{\bulk,\sigma}$ was state independent, $\langle K_{\bulk,\sigma}^X \rangle_{\rho}=\frac{c(X)}{\epsilon}$ and thus we recover a linear expression.\footnote{By state independent we mean in the smaller Hilbert space of bulk low energy excitations. This contribution would depend on the semiclassical background, since the RT surface and the metric will be different.}  $\langle K_{\bulk,\sigma} \rangle_{\rho}$ could in principle have state dependent divergences. State dependent divergences in the entropy were studied in \cite{Marolf:2016dob}, and they look like $\langle \int_{\partial R} {\cal O} \rangle_{\rho}$, which using the first law they can be mapped to a contribution to the modular Hamiltonian $\int_{\partial R} {\cal O}$ \cite{Jafferis:2015del}, which will lead to state dependent divergences in the modular Hamiltonian.  However, because our contribution to the entropy includes $\langle A_{\gen} \rangle$, $S_{\gen}$ will not have these divergences. In other words, $K_{\bulk,\sigma}+A_{\gen}$ does not have state dependent bulk divergences and we can just shift the possible term from $K_{\bulk,\sigma}$ to $A_{\gen}$ in a way that none of the terms will have state dependent divergences and  we get a clearly linear \er{1stterm}.

%%%%%%%%%%%%%%%%%%%%%%%%%%%%%%%%%%%%%%%%%%%%%%%%%%
\subsection{Modular extremality and the $G_N$ expansion}

One should think of the $G_N$ expansion of $ \dfrac{A^{X_{\sigma}(\rho)}}{4 G_N}+\langle K_{\bulk,\sigma}^{X_{\sigma}} \rangle$ as expanding around the classical extremal surface ${\cal K}_{X_{\ext}}^I(g_{cl,\rho})=0$. In terms of $\delta X_{\sigma}(\rho) \equiv  X_{\sigma}-X_{\ext}$, we can expand the area:
\begin{align}
 A^{X_{\sigma}(\rho)} = A^{X_{\ext}(\rho)}+\int dy f^{(1)}(g_{cl},y) \delta X_{\sigma}(y)^2+ \int dy f^{(2)}(g_{cl},y) \delta X_{\sigma}(y)^3+\cd.
\end{align}
And also the bulk modular Hamiltonian:
\begin{align}
\label{bulkmodhamexp}
\langle K_{\bulk,\sigma}^{X_{\sigma}} \rangle_{\rho} = \langle K_{\bulk,\sigma}^{X_{\ext}} \rangle_{\rho}+\int dy F^{(1)}_{K_{\sigma},\rho}[y] \delta X_{\sigma}(y)+\int dy dy' F^{(2)}_{K_{\sigma},\rho}[y,y'] \delta X_{\sigma}(y) \delta X_{\sigma}(y')+\cd
\end{align}
where the $F$'s are determined using modular perturbation theory, for example $F^{(1)}[y]$ is the RHS of  \nref{modext}. Modular extremality relates the two terms, schematically: $2 f^{(1)}(y) \delta X_{\sigma}+3 f^{(2)}(y) \delta X_{\sigma}(y)^2+\cd=- G_N (F^{(1)}[y]+2 \int dy' F^{(2)}[y,y'] \delta X_{\sigma}[y']+\cd)$, but there are no miraculous cancellations because the terms which are the same order in $G_N$ in the area and bulk modular Hamiltonian have different powers of $\delta X_{\sigma}$. Modular extremality does simplify the expression for the boundary modular Hamiltonian since it can be expressed purely in terms of only $\delta X_{\sigma}$ and $f$ (provided that we know $\delta X_{\sigma}$).  Of course, this expansion also applies for quantum extremal surfaces (when $\rho=\sigma$). 

From this expansion, one could require the relative entropy to be given by the bulk relative entropy of some surface, which is neither the modular nor quantum extremal surface. We could set up an equation
 \begin{equation}
 S_{\rel}(\rho|\sigma) \equiv \langle K_{\sigma} \rangle_{\rho}-\langle K_{\rho} \rangle_{\rho}= \langle A^{X_{\sigma}}-A^{X_{\rho}}+K^{X_{\sigma}}_{\bulk,\sigma}-K_{\bulk,\rho}^{X_{\rho}} \rangle_{\rho}=  \langle K^{X_{S}(\rho,\sigma)}_{\bulk,\sigma}-K_{\bulk,\rho}^{X_{S}(\rho,\sigma)} \rangle_{\rho} 
 \end{equation}
which should be solved order by order in $G_N$ by expanding the RHS using \nref{bulkmodhamexp} and solving for $X_{S}(\rho,\sigma)$. While it is clear that this can be done to leading order, we are not completely sure if there it has a solution to all orders. If that is true, it might be helpful to think about the interpretation of modular extremality: it relates variations of the area with variations of the modular Hamiltonian and this can be used to write the relative entropy as the bulk relative entropy in some $X_S$ surface. However, even if it is the case, it is clear that $X_S$ will be complicated and nonlinear in $\rho, \sigma$. 

%%%%%%%%%%%%%%%%%%%%%%%%%%%%%%%%%%%%%%%%%%%%%%%%%%
\subsection{$\langle K_{\bulk,\sigma} \rangle_{\rho}$ and local modular Hamiltonians}

At this point, even if we have a formal definition for this modular extremal surfaces, it would be nice to understand better what the different terms mean. 

 To compute $\langle K_{\bulk,\sigma} \rangle_{\rho}$, in gravity in $G_N$ perturbation theory, we have to account for three facts: the surface changes, the background metric changes, the quantum state changes. Only the latter is present in usual field theories. As we discussed before, the fact that the surface changes can be understood in terms of entanglement perturbation theory (and can be combined with the change in the area), and we are going to ignore this dependence in this section. Given that the background metric changes, we should think of the change in the state as a combination of a change in the matter fields plus a shift in the metric due to backreaction. We could deal with by deforming the path integral inserting an operator that changes the metric and this would give us a deformed modular Hamiltonian, as for shape deformations. However, given that the theory is gravitational there seems to be a more natural way to do it: we should think of the bulk modular Hamiltonian in terms of the $G_N$ expansion, to leading order it will be quadratic on the fields and then interactions will be present at higher orders.  Backreaction is easily introduces by just shifting the tadpole $g_{cl}$ which appears in the modular Hamiltonian, that is $K_{\bulk,\sigma}[g_{cl,\rho},h_{\rho}]\equiv K_{\bulk,\sigma}[g_{cl,\sigma},h_{\sigma}-(g_{\rho}-g_{\sigma}) ]$. This is just a shift of the variables, but the different expressions are useful when evaluated in the respective $g_{cl}$ state. 
 
As an example, we can consider $K_{\sigma}$, the modular Hamiltonian of a sphere $R$ in the vacuum and $\rho$ some state which varies by an $O(1)$ expectation value of the boundary stress tensor. In this case, the modular Hamiltonian is local:
\begin{equation}
\langle K_{\sigma} \rangle_{\rho}=\langle \int_R \xi.T \rangle_{\rho}.
\end{equation} 

When we have local modular Hamiltonian, we can use Wald's version of Gauss' law \cite{Wald:1993nt,Iyer:1994ys} (see also \cite{Jafferis:2015del}): 
\begin{equation}
 \langle \int_R \xi.T \rangle_{\rho=\sigma+\delta \rho}-\langle \int_R \xi.T \rangle_{\sigma}= {\cal E}_{\infty}(\delta g)=\int_{\Sigma_S} \xi_{\bulk}^t E_{tt,lin}(\delta g)+A^S_{lin}(\delta g)
\label{gauss} \end{equation}
 where $S$ is an arbitrary gauge-invariant surface that is well defined for the original and the perturbed state  (for example by picking a gauge where the surface stays at the same position). $\Sigma_S$ is the spacelike surface between the boundary region $R$ and the surface $S$. 
 
 Now, we can use \nref{gauss} to integrate in $\langle K_{\sigma} \rangle_{\rho}$ for $\rho$ perturbatively close to $\sigma$, to all orders in perturbation theory. The reason is simple, if we write $g_{\rho}=g_{\sigma}+\sum_k \lambda^k \delta^k g$, we have that $\langle \int_R \xi.T \rangle_{\rho}-\langle \int_R \xi.T \rangle_{\sigma}=\sum_k \lambda^k {\cal E}_{\infty}(\delta^k g)$ is linear in the metric (and $\rho$) and we can use the gravitational Gauss' law for each term individually.
 
Now, $E_{tt,lin}(\delta^k g)$ is nothing but the tadpole of equation \nref{Eingrav} (technically, \nref{Eingrav} referred to the $G_N$ expansion, but it of course applies to any other perturbative expansion) which we can morally think of as the stress tensor to that order. So, we can write the previous formula as:
\begin{equation}\label{gausslaw}
\langle K_{\sigma} \rangle_{\rho}-\langle K_{\sigma} \rangle_{\sigma}=\sum_k \lambda^k \int_{\Sigma_S} \xi. T^{(k)}_{grav}+A^{S}_{lin}(\delta^k g).
\end{equation}

 We expect that this can be used to write $K_{\text{bdy}}=K^{S}_{\bulk}+A^S$ for an arbitrary gauge invariant surface $S$, but this requires a careful analysis of boundary terms which we will not pursue further.\footnote{Although naively only the linearized area operator appears, the RHS of Einstein equations ($T_{grav}$) is the bulk modular Hamiltonian modulo boundary terms which turn the linearized area operator into the full area operator. One can see how this works to second order by carefully rewriting $T_{grav}$ as the canonical energy (bulk modular Hamiltonian) plus the quadratic area operator \cite{Jafferis:2015del}.} This means that modular extremality is not very helpful for local modular Hamiltonians. As the surface $S$, the most natural candidates  are classical extremal or modular extremal surfaces, but one could choose any other families of gauge invariant surfaces. It is clear from this discussion that we should think of the change in background in $K_{\bulk,\sigma}$ as simply shifting the tadpole from $g_{\sigma}$ to $g_{\rho}$.

Now, we would to connect the previous story with that of \cite{Lashkari:2016idm}. We can think of their setup in our terms as $\rho$ being a bulk coherent state on top of $\sigma$ with a semiclassical amplitude, schematically $|\Psi_{\rho}\rangle=e^{i \sqrt{\lambda /G_N} a^{\dagger}}|\Psi_{\sigma}\rangle$, with $a^{\dagger}$ the graviton creation operator. We can to work in the limit where the amplitude is large (so that the state is classical) but the states only change the metric perturbatively in $\lambda$. Since  $g_{\rho},g_{\sigma}$ correspond to the same saddle, we can apply our discussion. In this limit, even if in the entanglement entropy the area changes to order $G_N^{-1}$, the bulk entanglement entropy stays $O(G_N^{0})$, so we do not need quantum extremality. It is less clear if the modular extremal surface changes for coherent states, but we do not need it because of \nref{gausslaw}. We can instead consider the simpler case when $S$ is the extremal surface. In this case, since the bulk entanglement entropy is $O(G_N^0)$, but the bulk modular Hamiltonian is $O(G_N^{-1})$, we deduce that:
\begin{equation}
S_{\rel}(\rho|\sigma)=\Delta \langle K^{{S_{\ext}}}_{\bulk} \rangle+O(G_N^{0})
\end{equation}
where we used our expectation that $K_{\text{bdy}}=K^{S}_{\bulk}+A^S$ and for $S$ being the extremal surface the areas cancel in the relative entropy, they would not cancel for modular extremal surfaces.   In this way, it is very suggestive to think of the Hamiltonian of \cite{Lashkari:2016idm} as the bulk modular Hamiltonian in the entanglement wedge, in which case the positivity of relative entropy would be a consequence of the positivity of the bulk relative entropy. Again, modular extremality does not seem important in their case because in this symmetric situation, one can choose an arbitrary gauge invariant surface where to integrate the boundary modular Hamiltonian. Of course, to make full connection between \nref{gausslaw}, modular extremality and \cite{Lashkari:2016idm} more precise, one should understand better how the boundary terms and  $E_{lin}$ combine to give the bulk modular Hamiltonian to all orders.
 
More broadly, understanding if (classical) coherent states give an $O(1)$ shift to the position extremal surface when considering  modular extremality seems interesting, since quantum extremal surfaces can only shift the entangling surface by $O(G_N)$. This might give a simpler classical setup to compute the dual of the modular Hamiltonian.  For example, if we consider a coherent state of scalar fields, where $\langle \phi \rangle_{{\lambda}}=0+\sqrt{\lambda G_N^{-1}} \phi_{cl}$, we expect the modular extremal surface to shift by a classical $ \delta X^I(X,y) \propto {\lambda}^2 \int \frac{ds}{\sinh^2 (s/2+i \varepsilon)} T^{I r}_{s}$ when computing  $\langle K_{\text{bdy}}({\lambda}=0) \rangle_{{\lambda}}$ holographically. Of course, this is hard to do explicitly, because we have little control over modular Hamiltonians other than those which are local, where we can apply \nref{gausslaw}. 

%%%%%%%%%%%%%%%%%%%%%%%%%%%%%%%%%%%%%%%%%%%%%%%%%%
\section{Discussion}

In this paper, we have exploited the variational principle at the level of the replicated  path integral to derive the extremality of the entangling functional of higher derivatives, quantum extremality and modular extremality. This is done by thinking about the R\'{e}nyi entropies and taking the $n \rightarrow 1$ limit carefully.  This gives closure to the approach of \cite{Lewkowycz:2013nqa} which naturally gives the entanglement entropy functional but  makes it hard to derive the extremality condition for general gravitational theories and higher orders in $G_N$. This variational framework is also useful to generalize relation between the equations of motion and the first law for general states. 

We would like to close with some comments and future directions. 

As a general note, across this paper, we have assumed that the bulk saddles have replica symmetry. It would be nice if one could relax this or justify it better (see \cite{Faulkner:2013yia,Camps:2014voa} for some discussion about this ). 

%%%%%%%%%%%%%%%%%%%%%%%%%%%%%%%%%%%%%%%%%%%%%%%%%%
\subsubsection*{Higher-derivative gravity}

By working at integer $n>1$ and then taking the $n \rightarrow 1$ limit in higher-derivative theories of gravity, we have discussed how one should in principle determine the splitting terms of \cite{Miao:2014nxa}. These are determined by demanding that the gravitational action is finite. After fixing these terms, the only remaining freedom comes from changing the location of the surface, and this deformation keeps the action finite. 

In Appendix \ref{appendixhigher}, we have demonstrated in some nontrivial examples how the $n\to1$ limit of the Wald entropy at $n>1$ gives the gravitational entropy of \cite{Dong:2013qoa,Camps:2013zua}. While our approach is strongly suggestive that this is true generally, it would be useful to work it out explicitly for more general examples. 

%%%%%%%%%%%%%%%%%%%%%%%%%%%%%%%%%%%%%%%%%%%%%%%%%%
\subsubsection*{The equations of motion} 

About the equations of motion, it would be nice to understand better if by varying the regions that in consideration, one can derive the local equations of motion from the integrated equations of motion. Note that, in contrast with \cite{Faulkner:2013ica}, the equations are integrated over one more dimension because of the lack of symmetry.

In order to derive the equations of motion from the first law of entanglement of \cite{Blanco:2013joa,Wong:2013gua}, one has to understand the modular Hamiltonian. In general it is complicated, yet its variations are well defined in terms of analytically continued one point functions in the replicated theory. We expect that, in the absence of a more explicit expression for the boundary Hamiltonian, the only way in which one can obtain the equations of motion from the first law is by using the replica trick via the procedure described in Section \ref{EOM}. 

 Of course, there are other ways in which one could try to get the equations of motion from the RT formula. An alternative option pursued by \cite{Faulkner:2014jva,Faulkner:2017tkh} is to show that the boundary expression for the relative entropy around the vacuum for a spherical region matches the expression for bulk relative entropy. The bulk and boundary relative entropies differ off-shell by an integral of the equations of motion and thus one can derive the backreacted equations of motion from the equality of these two quantities. More generally, one might be able to use similar ideas to the ones that we described combined with modular perturbation theory to generalize this approach to other surfaces and states. 

%%%%%%%%%%%%%%%%%%%%%%%%%%%%%%%%%%%%%%%%%%%%%%%%%%
\subsubsection*{Entanglement entropy of gravitons}

We defined the entanglement entropy of gravitons by analytically continuing the finite $n-1$ partition function. Technically speaking, only $S_{\gen}$ is well defined, since the separation into two terms is ambiguous: it depends on the details of how the boundary is inserted. This ambiguity is related with the choice of center of \cite{Casini:2013rba}. It would be nice to understand better the graviton entanglement entropy from a Hilbert space perspective, along the lines of \cite{Donnelly:2014fua,Donnelly:2015hxa,Jafferis:2015del}.  It would be interesting to carry out the perturbation theory described in Section \ref{definitionqext} to define the entanglement entropy of gravitons beyond the extremal surfaces in $G_N$ perturbation theory. 

%%%%%%%%%%%%%%%%%%%%%%%%%%%%%%%%%%%%%%%%%%%%%%%%%%
\subsubsection*{Local modular Hamiltonians and modular extremality}

We have also given an argument of how one can in principle think of the results of \cite{Lashkari:2016idm} in terms of bulk relative entropy. Of course, it would be nice to understand this more precisely, by being careful about the boundary terms in the graviton modular Hamiltonian to higher orders. 

Modular extremality does not seem necessary when the modular Hamiltonian is local, since there we can just use Gauss' law to integrate in the energy at infinity. It seems hard yet very interesting to understand explicitly some examples of modular extremality for modular Hamiltonians which are non-local. In contrast with quantum extremality , we expect the modular extremal surface to be different from the extremal surface for deformations which are classical (coherent states).  

%%%%%%%%%%%%%%%%%%%%%%%%%%%%%%%%%%%%%%%%%%%%%%%%%%
\subsubsection*{Modular flow and bulk reconstruction}

To leading order in $G_N$, the commutator between a properly dressed local operator at a point $Z$ in the entanglement wedge and the modular Hamiltonian is given by the commutator with the bulk modular Hamiltonian. This was used in \cite{Dong:2016eik} to show that one can reconstruct operators in the entanglement wedge in terms of the boundary subregion and more recently, it was used in \cite{Faulkner:2017vdd} to derive a boundary expression of the bulk operators.
 
Furthermore, \cite{Dong:2016eik} showed that if $ \rho_{\bulk}=\sigma_{\bulk} \rightarrow \rho=\sigma $, which is clearly true from \nref{rem}, then one can also reconstruct  operators deep inside the entanglement wedge. As has been argued recently \cite{Cotler:2017erl}, the analysis of \cite{Almheiri:2014lwa,Dong:2016eik} is stable under $G_N$ perturbations and we expect that our discussion can help find the explicit bulk to boundary mapping in the presence of backreaction. Because of the previous, we do not expect the approach of \cite{Faulkner:2017vdd} to break down when $G_N$ corrections are considered.  To next order, it seems like the correction to the difference between modular flows is determined by the shift in the surface:
\begin{equation}
[K_{\sigma},\Phi(Z)]=[K_{\bulk,\sigma},\Phi(Z)]+G_N \int_{RT} dy [\delta X(y)^2,\Phi(Z)].
\end{equation}
 
We leave for future work understanding this contribution to the commutator, but we expect that by carefully understanding the previous one can generalize \cite{Faulkner:2017vdd} to higher orders in $G_N$.  

%%%%%%%%%%%%%%%%%%%%%%%%%%%%%%%%%%%%%%%%%%%%%%%%%%
\section*{Acknowledgments}
We would like to thank Eric D'Hoker, Tom Faulkner, Daniel Harlow, Nima Lashkari, Juan Maldacena, Onkar Parrikar, Mukund Rangamani, Douglas Stanford, and Aron Wall for useful discussions.  X.D. was supported in part by the National Science Foundation under Grant No.~PHY-1606531, by the Department of Energy under Grant No.~DE-SC0009988 and, by the Martin A. and Helen Chooljian Founders' Circle Membership at the Institute for Advanced Study.  A.L. acknowledges support from the Simons Foundation through the It from Qubit collaboration, as well as the support of a Myhrvold-Havranek Innovative Thinking Fellowship. A.L. would also like to thank the Department of Physics and Astronomy at the University of Pennsylvania for hospitality during the development of this work. 

\appendix
%%%%%%%%%%%%%%%%%%%%%%%%%%%%%%%%%%%%%%%%%%%%%%%%%%
\se{Dilaton gravity with higher derivative interactions}\la{appendixhigher}

In this appendix, we study the gravitational entropy in toy models of higher derivative gravity: 2d dilaton gravity coupled to matter fields with higher derivative interactions.  These theories can arise from dimensional reduction of higher derivative gravity in more than two dimensions.  We demonstrate how to solve the ``splitting problem'' and calculate the entropy functional $A_\gen$ in these toy models.  Furthermore, we verify \er{rth} and \er{Agen1} by showing directly from the equations of motion that the entropy is obtained by extremizing $A_\gen$, and its extremal value agrees with the $n\to1$ limit of the Wald entropy.

Throughout this appendix, we define $\e\eq n-1$ and adopt a complex coordinate system $(z,\zb)$ on $M_n$ such that the metric is in the conformal gauge
\be
ds^2 = e^{2\y(z,\zb)} dz d\zb
\ee
and the origin is the $\bZ_n$ fixed point.  The $\bZ_n$ symmetry acts as a discrete rotation $z\to z e^{2\pi i/n}$.

We will study solutions of the equations of motion for $\y$, the dilaton $\p$, and additional matter fields.  At $n=1$, these fields have regular Taylor expansions around $z=0$.  For example, we have
\be\la{exp}
\p(z,\zb) \Big|_{n=1} = \mr\p +\mr\p_z z +\mr\p_{\zb} \zb +\fr{1}{2} \mr\p_{zz} z^2 +\fr{1}{2}\mr\p_{\zb\zb} \zb^2 +\mr\p_{z\zb} z\zb +\cd
\ee
for the dilaton.  Away from $n=1$, such expansions become much more complicated.  Near $n\ap1$, we may generally expand the dilaton as
\bm\la{nexp}
\p(z,\zb)=\p_0 +\p_1 (z\zb)^\e +\p_2 (z\zb)^{2\e} +\cd +\lt\{z^{1+\e} \[\p_{z,0} +\p_{z,1} (z\zb)^\e +\cd\] + c.c.\rt\}\\
+\lt\{\fr{1}{2} z^{2(1+\e)} \[\p_{zz,0} +\p_{zz,1} (z\zb)^\e +\cd\] +c.c.\rt\} +z\zb \[\p_{z\zb,0} +\p_{z\zb,1} (z\zb)^\e +\cd\] +\cd
\em
and similarly for other fields.  Here $c.c.$ denotes complex conjugate.  As we go away from $n=1$, each term in the expansion \er{exp} ``splits'' into a Taylor expansion in $(z\zb)^\e$.  Continuity at $n=1$ therefore requires the following matching conditions:
\ba\la{psp}
\mr\p &= \p_0 +\p_1 +\p_2 +\cd,\\
\mr\p_\m &= \p_{\m,0} +\p_{\m,1} +\p_{\m,2} +\cd,\\\la{pmnsp}
\mr\p_{\m\n} &= \p_{\m\n,0} +\p_{\m\n,1} +\p_{\m\n,2} +\cd,
\ea
and their higher-order analogues.  Here $\m=z,\zb$, and we have only kept zeroth-order terms in $\e$ in coefficients such as $\p_m$ and $\p_{\m,m}$.  Higher-order terms in $\e$ are negligible for the purpose of calculating the von Neumann entropy in our examples.

The gravitational entropy $A_\gen$ can be calculated as in \cite{Dong:2013qoa}, but the result would depend on how the $n=1$ coefficients split into $n\neq 1$ coefficients in \er{psp}--\er{pmnsp}.  On the other hand, $A_\gen$ should depend only on the $n=1$ solution \er{exp} in order to be a useful entropy functional.  This is the ``splitting problem.''  As we will demonstrate explicitly below, the solution to this problem is that the equations of motion near $n\ap 1$ are sufficient to fix the split of coefficients in \er{psp}--\er{pmnsp}, at least to the extent of allowing us to write $A_\gen$ in terms of the $n=1$ coefficients appearing in \er{exp}.

%%%%%%%%%%%%%%%%%%%%%%%%%%%%%%%%%%%%%%%%%%%%%%%%%%
\sse{One matter field}

Let us first consider the following theory of dilaton gravity coupled with a single scalar field $\s$ with higher derivative interaction:
\be\la{aom}
I = -\fr{1}{2} \int d^2x \sqrt{g} \[\p R -(\na\s)^2 +\l \na_\m\na_\n\s \na^\m\na^\n\s\].
\ee
The equation of motion for the metric is
\bm
\fr{1}{\sqrt{g}} \fr{\d I}{\d g^{\m\n}} = g_{\m\n} \[-\fr{1}{2} \na^2\p -\fr{1}{4}(\na\s)^2 +\fr{\l}{4} \na_\r\na_\s\s \na^\r\na^\s\s\] +\fr{1}{2} \na_\m\na_\n\p +\fr{1}{2} \na_\m\s \na_\n\s \\
+\fr{\l}{2} \(\na_\m\s \na^2 \na_\n\s +\na_\n\s \na^2 \na_\m\s -\na^2\s \na_\m\na_\n\s -\na_\r\s \na^\r\na_\m\na_\n\s\) = 0,
\em
whereas the equations of motion for the dilaton $\p$ and the scalar $\s$ are
\ba\la{peom}
-\fr{1}{\sqrt{g}} \fr{\d I}{\d\p} &= \fr{1}{2}R=0,\\
-\fr{1}{\sqrt{g}} \fr{\d I}{\d\s} &= \na^2\s +\l \na_\m\na_\n \na^\m\na^\n\s=0.
\ea
Using \er{peom} we find a flat space with the conformal factor $\y=0$, greatly simplifying the other equations.  If we want, we could get an AdS solution instead by replacing $R$ with $R+2$ in \er{aom}; this leads to $\y= -\log\big(1-\fr{z\zb}{4}\big)$ but our conclusion is largely unaffected.

Solving the other equations of motion near $n\ap 1$, we find
\ba
&\s_{m>0}=0,\qu
\s_{\m,m>0}=0,\qu
\s_{\m\n,m>0}=0,\\
&\p_1 = 2\l \s_{z,0} \s_{\zb,0},\qu
\p_{m>1}=0,\\
&\p_{z,0}=2\l \s_{z,0} \s_{z\zb,0},\qu
\p_{z,1}=2\l \s_{\zb,0} \s_{zz,0},\qu
\p_{z,m>1}=0.
\ea
This holds for arbitrary $\l$ and constrains how the coefficients split in \er{psp}--\er{pmnsp}:
\ba\la{ssps}
&\s_0 = \mr\s,\qu
\s_{\m,0} = \mr\s_\m,\qu
\s_{\m\n,0} = \mr\s_{\m\n},\qu
\p_0 = \mr\p - 2\l \mr\s_z \mr\s_{\zb},\\\la{szs}
&\mr\p_z = 2\l \(\mr\s_z \mr\s_{z\zb} + \mr\s_{\zb} \mr\s_{zz}\).
\ea
Let us make two comments before continuing.  First, these relations are uniquely determined from a local analysis of the equations of motion near a small conical defect in the quotient space $\hat M_n$, and are universal in the sense that they do not depend on whatever boundary conditions we impose at the asymptotic boundary of spacetime.  The reason for this is that these relations arise from setting to zero the most singular terms in the equations of motion expanded around $z=0$.  This is a good feature because the entropy functional $A_\gen$ should only depend on local geometric quantities once we fix the gravitational action.  Our second comment is that the split of $\mr\s_z$ (and $\mr\s_{\zb}$) is over-constrained as shown in \er{szs}, but we will see that this is a feature not a bug.

The gravitational entropy can be easily calculated as in \cite{Dong:2013qoa}:
\be
A_\gen = 2\pi (\p_0+\p_1) -4\pi\l \s_{z,0} \s_{\zb,0}.
\ee
As promised, this entropy functional can be rewritten\footnote{This rewriting only uses the most singular part of the equations of motion expanded around $z=0$, and is valid even after a $\d g_n$ variation as long as it is regular as defined in Section \re{secbt}.} in terms of fields and their derivatives at $n=1$:
\be\la{agps}
A_\gen = S_{\Wald} + S_{\anomaly},\qu
S_{\Wald} = 2\pi\mr\p,\qu
S_{\anomaly} = -4\pi\l \mr\s_z \mr\s_{\zb}.
\ee
Moreover, it agrees with the $n\to1^+$ limit of the Wald entropy
\be
\lim_{n\to1^+} S_{\Wald}(g_n) = 2\pi \p_0
\ee
which is identical to \er{agps} after using \er{ssps}.  It is worth noting that in taking the above limit we need to calculate the Wald entropy at $n>1$, and $\p_1$ does not contribute to this.  Therefore, the Wald entropy has a discontinuity of the amount $2\pi \p_1$ at $n=1$, which is precisely compensated by $S_{\anomaly}$ in \er{agps}.

We satisfy the extremality condition $\pa_\m A_{\gen}=0$ because it reduces to
\be
\pa_z A_{\gen} = \pa_z (2\pi\mr\p -4\pi\l \mr\s_z \mr\s_{\zb}) = 2\pi\[\mr\p_z -2\l \(\mr\s_z \mr\s_{z\zb} + \mr\s_{\zb} \mr\s_{zz}\)\]
\ee
which vanishes due to the extra constraint \er{szs}.

%%%%%%%%%%%%%%%%%%%%%%%%%%%%%%%%%%%%%%%%%%%%%%%%%%
\sse{Two matter fields}

The previous example may seem too simple for experts, so let us now study a more complicated theory of dilaton gravity coupled with two scalar fields $\s$ and $\w$ with higher derivative interaction:
\be\la{itm}
I = -\fr{1}{2} \int d^2x \sqrt{g} \[\p R -(\na\s)^2 -(\na\w)^2 +\l \w \na_\m\na_\n\s \na^\m\na^\n\s\].
\ee
The equation of motion for the metric is
\bm\la{eomm}
\fr{1}{\sqrt{g}} \fr{\d I}{\d g^{\m\n}} = g_{\m\n} \[-\fr{1}{2} \na^2\p -\fr{1}{4}(\na\s)^2 -\fr{1}{4}(\na\w)^2 +\fr{\l}{4} \w \na_\r\na_\s\s \na^\r\na^\s\s\]\\
+\fr{1}{2} \na_\m\na_\n\p +\fr{1}{2} \na_\m\s \na_\n\s +\fr{1}{2} \na_\m\w \na_\n\w \\
+\fr{\l}{2} \Big\{\big[ \na_\m\s \na^\r (\w \na_\r \na_\n\s) +(\m\lra\n)\big] -\na^\r (\w \na_\r\s \na_\m \na_\n\s) \Big\} = 0,
\em
whereas the equations of motion for the dilaton $\p$ and the other scalars $\s$, $\w$ are
\ba
-\fr{1}{\sqrt{g}} \fr{\d I}{\d\p} &= \fr{1}{2}R =0,\\
-\fr{1}{\sqrt{g}} \fr{\d I}{\d\s} &= \na^2\s +\l \na_\m\na_\n (\w \na^\m\na^\n\s) =0,\\
-\fr{1}{\sqrt{g}} \fr{\d I}{\d\w} &= \na^2\w +\fr{\l}{2} \na_\m\na_\n\s \na^\m\na^\n\s =0.
\ea
Again we find a flat space with the conformal factor $\y=0$.

It is difficult to solve the other equations of motion for arbitrary $\l$, so we will work perturbatively in $\l$ and write the solution as
\be
\p = \p^{(0)} + \l \p^{(1)} + \l^2 \p^{(2)} +\cd
\ee
with similar expansions for other fields.

At the zeroth order in $\l$, we find the familiar case of dilaton gravity without any higher derivative interaction:
\ba
&\p_{m>0}^{(0)} =0,\qu
\p_{\m,m\ge0}^{(0)} =0,\\
&\s_{m>0}^{(0)} =0,\qu
\s_{\m,m>0}^{(0)} =0,\qu
\s_{z\zb,m\ge0}^{(0)} =0,\\
&\w_{m>0}^{(0)} =0,\qu
\w_{\m,m>0}^{(0)} =0,\qu
\w_{z\zb,m\ge0}^{(0)} =0.
\ea
At the linear order in $\l$, we find
\ba
&\p_1^{(1)} =2 \w_0^{(0)} \s_{z,0}^{(0)} \s_{\zb,0}^{(0)},\qu
\p_{m>1}^{(1)} =0,\\
&\p_{z,0}^{(1)} =0,\qu
\p_{z,1}^{(1)} =2 \s_{\zb,0}^{(0)} \[\w_0^{(0)} \s_{zz,0}^{(0)} +\w_{z,0}^{(0)} \s_{z,0}^{(0)}\],\qu
\p_{z,m>1}^{(1)} =0,\\
&\s_1^{(1)} = -\[\w_{z,0}^{(0)} \s_{\zb,0}^{(0)} + \w_{\zb,0}^{(0)} \s_{z,0}^{(0)}\],\qu
\s_{m>1}^{(1)} =0,\\
&\s_{z,1}^{(1)} =-\[\w_{zz,0}^{(0)} \s_{\zb,0}^{(0)} + \s_{zz,0}^{(0)} \w_{\zb,0}^{(0)}\],\qu
\s_{z,m>1}^{(1)} =0,\\
&\w_1^{(1)} = -\s_{z,0}^{(0)} \s_{\zb,0}^{(0)},\qu
\w_{m>1}^{(1)} =0,\qu
\w_{z,1}^{(1)} =-\s_{\zb,0}^{(0)} \s_{zz,0}^{(0)},\qu
\w_{z,m>1}^{(1)} =0.
\ea
At the second order in $\l$, all we need to find is
\ba
\p_1^{(2)} &= 2 \w_0^{(0)} \[\s_{z,0}^{(0)} \s_{\zb,0}^{(1)} +c.c.\] -2\w_0^{(1)} \w_1^{(1)},\\
\p_2^{(2)} &= \fr{1}{2} \[\s_1^{(1)}\]^2 -\fr{3}{2} \[\w_1^{(1)}\]^2 +2 \w_0^{(0)} \[\s_{z,0}^{(0)} \s_{\zb,1}^{(1)} +c.c.\],\qu
\p_{m>2}^{(2)} =0,
\ea

From these results we can determine the gravitational entropy as in \cite{Dong:2013qoa}.  Let us find the contribution to $A_\gen$ from each term in the action \er{itm}.  We will work to second order in $\l$.  The contribution of the $\p R$ term is
\be
2\pi \[\p_0^{(0)} +\l\(\p_0^{(1)} +\p_1^{(1)}\) +\l^2 \(\p_0^{(2)} +\p_1^{(2)} +\p_2^{(2)}\)\].
\ee
From the $(\na\s)^2$ term we get
\be
\pi\l^2 \[\s_1^{(1)}\]^2,
\ee
whereas the contribution of the $(\na\w)^2$ term is
\be
\pi\l^2 \[\w_1^{(1)}\]^2.
\ee
From the $\l \w \na_\m\na_\n\s \na^\m\na^\n\s$ term we get the contribution
\bm
-4\pi\l \[\w_0^{(0)}+ \l\(\w_0^{(1)} +\fr{1}{2}\w_1^{(1)}\)\] \[\s_{z,0}^{(0)} +\l\(\s_{z,0}^{(1)} +\s_{z,1}^{(1)}\)\] \[\s_{\zb,0}^{(0)} +\l\(\s_{\zb,0}^{(1)} +\s_{\zb,1}^{(1)}\)\] \\
+2\pi \l^2 \s_1^{(1)} \[\w_{z,0}^{(0)} \s_{\zb,0}^{(0)} +\w_{\zb,0}^{(0)} \s_{z,0}^{(0)}\].
\em

Combining these four contributions we get the gravitational entropy
\be\la{agso}
A_\gen = A_\gen^{(0)} +\l A_\gen^{(1)} +\l^2 A_\gen^{(2)} +\cd
\ee
where
\ba
A_\gen^{(0)} &= 2\pi \p_0^{(0)},\\
A_\gen^{(1)} &= 2\pi \[\p_0^{(1)}+\p_1^{(1)}\] -4\pi \w_0^{(0)} \s_{z,0}^{(0)} \s_{\zb,0}^{(0)},\\
A_\gen^{(2)} &= 2\pi \[\p_0^{(2)} +\p_1^{(2)} +\p_2^{(2)}\] +3\pi \[\w_1^{(1)}\]^2 -\pi \[\s_1^{(1)}\]^2\nonumber\\
&\qqu\qqu\qqu+4\pi \w_0^{(1)} \w_1^{(1)} -4\pi \w_0^{(0)} \[\s_{z,0}^{(0)} \(\s_{\zb,0}^{(1)} + \s_{\zb,1}^{(1)}\) +c.c.\].
\ea
Again, this entropy function can be rewritten in terms of fields and their derivatives at $n=1$.  To second order in $\l$ we find
\be\la{agc}
A_\gen = 2\pi \mr\p -4\pi \l \mr\w \mr\s_z \mr\s_{\zb} -\pi \l^2 \[\mr\s_z^2 \mr\s_{\zb}^2 +\(\mr\w_z \mr\s_{\zb} +\mr\w_{\zb} \mr\s_z\)^2\] +O(\l^3).
\ee
This also agrees with the $n\to1^+$ limit of the Wald entropy
\be
\lim_{n\to1^+} S_{\Wald}(g_n) = 2\pi \[\p_0^{(0)} +\l \p_0^{(1)} +\l^2 \p_0^{(2)} \] +O(\l^3)
\ee
which can easily be shown to be identical to \er{agso}.

It is worth noting that if we forgot about splitting and proceeded na\"{\i}vely, we would miss the $\l^2$ term in \er{agc}.  Therefore, this example shows that we cannot in general forget about splitting in calculating the gravitational entropy.

It is possible to check the extremality condition $\pa_\m A_\gen=0$ by working out the relevant part of the $zz$ component of \er{eomm} to second order in $\l.$

%%%%%%%%%%%%%%%%%%%%%%%%%%%%%%%%%%%%%%%%%%%%%%%%%%
\section{Polyakov action}\label{appendixPoly}

A toy model to understand these issues would be to consider 2d dilaton gravity in the presence of $m$ quantum scalar fields \cite{Callan:1992rs,Russo:1992ax}
\begin{equation}
I=\frac{1}{2 \pi} \int dx^2 \sqrt{g} \left[ e^{-2 \phi} (R+4(\partial \phi)^2+4 \lambda^2) \right ]-\frac{m \hbar}{96 \pi} \int R \nabla^{-1} R.
\end{equation}

In the limit of large $m$, one can analyze the theory at  finite $N=m \hbar$. The second term can be thought of as $\int (\partial \eta)^2-2 \eta R$, with $\nabla \eta =R$. This expression suggests that $S_{\Wald}=\frac{N}{12} \eta_0$. This might seem too quick, but \cite{Myers:1994sg} showed using the Noether charge methods that $S_{\Wald}=\frac{N }{12} \eta_0$, so that the total entropy is \begin{equation}
S_{total}=2 e^{-2 \phi_0}+\frac{N}{12} \eta_0
=2 e^{-2 \phi_0}+\frac{N}{6} \rho_0
\end{equation}  
where $\eta_0$, which is non-local in general, was expressed in terms of the metric in conformal gauge, $ds^2=e^{2 \rho} dz d \bar{z}$.  The quantum extremality condition would be $-4 e^{-2 \phi_0} \partial \phi_0+\frac{N}{6} \partial \rho_0=0$. 

The equations of motion are \cite{Russo:1992ax}:
\begin{equation}
0=e^{-2\phi} (4 \partial \rho \partial \phi+2 \partial^2 \phi)-\frac{N}{12} (\partial \rho \partial \rho+\partial^2 \rho)
\end{equation}
and similarly for $\bar{\partial}$. 

Now, the question is whether given some metric $\rho$, the equations of motion can be solved if one adds a small conical singularity $\delta_n \rho=(n-1) \log z \bar{z}$. If $N=0$, then it was shown \cite{Lewkowycz:2013nqa} that a $\delta_n \phi$ change cannot cancel the singularity of $\partial \delta \rho = \frac{(n-1)}{z}$, so one concludes that $\partial \phi=0$. 

In the presence of $N$, there will be two kind of terms linear in $\delta_n$:
\begin{equation}
 \partial \delta_n \rho (4 \partial \phi e^{-2 \phi}-\frac{N}{6} \partial \rho)+\left [\delta_n (e^{-2 \phi} \partial \phi) \partial \rho+2 \delta_n(e^{-2 \phi} \partial^2  \phi)-\frac{N}{12} \partial^2 \delta_n \rho \right ]=0
\end{equation}
with $\delta_n \rho=(n-1) \log z \bar{z}$. If we consider $\phi=\rho=0$, then the equation is solved by setting $\delta_n \phi=\frac{N}{24} \delta_n \rho$. For a non-trivial background, one can cancel the $\frac{n-1}{z^2}$ between brackets by picking an appropriate $\delta_n \phi$. This then results in the condition $(4 \partial \phi e^{-2 \phi}-\frac{N}{6} \partial \rho)=0$ which is the quantum extremality condition. 

Naively, it seems non trivial that one would get the quantum extremality condition because the gravitational action is non-local. However, in this particular case, after adding an extra field the action becomes local and thus the usual arguments apply. 

\bibliographystyle{JHEP}
\bibliography{bibliography}

\providecommand{\href}[2]{#2}\begingroup\raggedright\begin{thebibliography}{10}

\bibitem{Bekenstein:1973ur}
J.~D. Bekenstein, \emph{{Black holes and entropy}},
  \href{https://doi.org/10.1103/PhysRevD.7.2333}{\emph{Phys.Rev.} {\bfseries
  D7} (1973) 2333}.

\bibitem{Bardeen:1973gs}
J.~M. Bardeen, B.~Carter and S.~Hawking, \emph{{The Four laws of black hole
  mechanics}},
  \href{https://doi.org/10.1007/BF01645742}{\emph{Commun.Math.Phys.} {\bfseries
  31} (1973) 161}.

\bibitem{Hawking:1974sw}
S.~Hawking, \emph{{Particle Creation by Black Holes}},
  \href{https://doi.org/10.1007/BF02345020}{\emph{Commun.Math.Phys.} {\bfseries
  43} (1975) 199}.

\bibitem{Maldacena:1997re}
J.~M. Maldacena, \emph{{The Large N limit of superconformal field theories and
  supergravity}}, {\emph{Adv.Theor.Math.Phys.} {\bfseries 2} (1998) 231}
  [\href{https://arxiv.org/abs/hep-th/9711200}{{\ttfamily hep-th/9711200}}].

\bibitem{Gubser:1998bc}
S.~Gubser, I.~R. Klebanov and A.~M. Polyakov, \emph{{Gauge theory correlators
  from noncritical string theory}},
  \href{https://doi.org/10.1016/S0370-2693(98)00377-3}{\emph{Phys.Lett.}
  {\bfseries B428} (1998) 105}
  [\href{https://arxiv.org/abs/hep-th/9802109}{{\ttfamily hep-th/9802109}}].

\bibitem{Witten:1998qj}
E.~Witten, \emph{{Anti-de Sitter space and holography}},
  {\emph{Adv.Theor.Math.Phys.} {\bfseries 2} (1998) 253}
  [\href{https://arxiv.org/abs/hep-th/9802150}{{\ttfamily hep-th/9802150}}].

\bibitem{Ryu:2006bv}
S.~Ryu and T.~Takayanagi, \emph{{Holographic derivation of entanglement entropy
  from AdS/CFT}},
  \href{https://doi.org/10.1103/PhysRevLett.96.181602}{\emph{Phys.Rev.Lett.}
  {\bfseries 96} (2006) 181602}
  [\href{https://arxiv.org/abs/hep-th/0603001}{{\ttfamily hep-th/0603001}}].

\bibitem{Ryu:2006ef}
S.~Ryu and T.~Takayanagi, \emph{{Aspects of Holographic Entanglement Entropy}},
  \href{https://doi.org/10.1088/1126-6708/2006/08/045}{\emph{JHEP} {\bfseries
  08} (2006) 045} [\href{https://arxiv.org/abs/hep-th/0605073}{{\ttfamily
  hep-th/0605073}}].

\bibitem{Lewkowycz:2013nqa}
A.~Lewkowycz and J.~Maldacena, \emph{{Generalized gravitational entropy}},
  \href{https://doi.org/10.1007/JHEP08(2013)090}{\emph{JHEP} {\bfseries 08}
  (2013) 090} [\href{https://arxiv.org/abs/1304.4926}{{\ttfamily 1304.4926}}].

\bibitem{Hubeny:2007xt}
V.~E. Hubeny, M.~Rangamani and T.~Takayanagi, \emph{{A Covariant holographic
  entanglement entropy proposal}},
  \href{https://doi.org/10.1088/1126-6708/2007/07/062}{\emph{JHEP} {\bfseries
  07} (2007) 062} [\href{https://arxiv.org/abs/0705.0016}{{\ttfamily
  0705.0016}}].

\bibitem{Dong:2016hjy}
X.~Dong, A.~Lewkowycz and M.~Rangamani, \emph{{Deriving covariant holographic
  entanglement}}, \href{https://doi.org/10.1007/JHEP11(2016)028}{\emph{JHEP}
  {\bfseries 11} (2016) 028}
  [\href{https://arxiv.org/abs/1607.07506}{{\ttfamily 1607.07506}}].

\bibitem{Dong:2013qoa}
X.~Dong, \emph{{Holographic Entanglement Entropy for General Higher Derivative
  Gravity}}, \href{https://doi.org/10.1007/JHEP01(2014)044}{\emph{JHEP}
  {\bfseries 01} (2014) 044} [\href{https://arxiv.org/abs/1310.5713}{{\ttfamily
  1310.5713}}].

\bibitem{Camps:2013zua}
J.~Camps, \emph{{Generalized entropy and higher derivative Gravity}},
  \href{https://doi.org/10.1007/JHEP03(2014)070}{\emph{JHEP} {\bfseries 03}
  (2014) 070} [\href{https://arxiv.org/abs/1310.6659}{{\ttfamily 1310.6659}}].

\bibitem{Blanco:2013joa}
D.~D. Blanco, H.~Casini, L.-Y. Hung and R.~C. Myers, \emph{{Relative Entropy
  and Holography}}, \href{https://doi.org/10.1007/JHEP08(2013)060}{\emph{JHEP}
  {\bfseries 08} (2013) 060} [\href{https://arxiv.org/abs/1305.3182}{{\ttfamily
  1305.3182}}].

\bibitem{Wong:2013gua}
G.~Wong, I.~Klich, L.~A. Pando~Zayas and D.~Vaman, \emph{{Entanglement
  Temperature and Entanglement Entropy of Excited States}},
  \href{https://doi.org/10.1007/JHEP12(2013)020}{\emph{JHEP} {\bfseries 12}
  (2013) 020} [\href{https://arxiv.org/abs/1305.3291}{{\ttfamily 1305.3291}}].

\bibitem{Lashkari:2013koa}
N.~Lashkari, M.~B. McDermott and M.~Van~Raamsdonk, \emph{{Gravitational
  dynamics from entanglement 'thermodynamics'}},
  \href{https://doi.org/10.1007/JHEP04(2014)195}{\emph{JHEP} {\bfseries 04}
  (2014) 195} [\href{https://arxiv.org/abs/1308.3716}{{\ttfamily 1308.3716}}].

\bibitem{Faulkner:2013ica}
T.~Faulkner, M.~Guica, T.~Hartman, R.~C. Myers and M.~Van~Raamsdonk,
  \emph{{Gravitation from Entanglement in Holographic CFTs}},
  \href{https://doi.org/10.1007/JHEP03(2014)051}{\emph{JHEP} {\bfseries 03}
  (2014) 051} [\href{https://arxiv.org/abs/1312.7856}{{\ttfamily 1312.7856}}].

\bibitem{Czech:2012bh}
B.~Czech, J.~L. Karczmarek, F.~Nogueira and M.~Van~Raamsdonk, \emph{{The
  Gravity Dual of a Density Matrix}},
  \href{https://doi.org/10.1088/0264-9381/29/15/155009}{\emph{Class. Quant.
  Grav.} {\bfseries 29} (2012) 155009}
  [\href{https://arxiv.org/abs/1204.1330}{{\ttfamily 1204.1330}}].

\bibitem{Wall:2012uf}
A.~C. Wall, \emph{{Maximin Surfaces, and the Strong Subadditivity of the
  Covariant Holographic Entanglement Entropy}},
  \href{https://doi.org/10.1088/0264-9381/31/22/225007}{\emph{Class. Quant.
  Grav.} {\bfseries 31} (2014) 225007}
  [\href{https://arxiv.org/abs/1211.3494}{{\ttfamily 1211.3494}}].

\bibitem{Headrick:2014cta}
M.~Headrick, V.~E. Hubeny, A.~Lawrence and M.~Rangamani, \emph{{Causality \&
  holographic entanglement entropy}},
  \href{https://doi.org/10.1007/JHEP12(2014)162}{\emph{JHEP} {\bfseries 12}
  (2014) 162} [\href{https://arxiv.org/abs/1408.6300}{{\ttfamily 1408.6300}}].

\bibitem{Faulkner:2013ana}
T.~Faulkner, A.~Lewkowycz and J.~Maldacena, \emph{{Quantum corrections to
  holographic entanglement entropy}},
  \href{https://doi.org/10.1007/JHEP11(2013)074}{\emph{JHEP} {\bfseries 11}
  (2013) 074} [\href{https://arxiv.org/abs/1307.2892}{{\ttfamily 1307.2892}}].

\bibitem{Barrella:2013wja}
T.~Barrella, X.~Dong, S.~A. Hartnoll and V.~L. Martin, \emph{{Holographic
  entanglement beyond classical gravity}},
  \href{https://doi.org/10.1007/JHEP09(2013)109}{\emph{JHEP} {\bfseries 09}
  (2013) 109} [\href{https://arxiv.org/abs/1306.4682}{{\ttfamily 1306.4682}}].

\bibitem{Engelhardt:2014gca}
N.~Engelhardt and A.~C. Wall, \emph{{Quantum Extremal Surfaces: Holographic
  Entanglement Entropy beyond the Classical Regime}},
  \href{https://doi.org/10.1007/JHEP01(2015)073}{\emph{JHEP} {\bfseries 01}
  (2015) 073} [\href{https://arxiv.org/abs/1408.3203}{{\ttfamily 1408.3203}}].

\bibitem{Jafferis:2015del}
D.~L. Jafferis, A.~Lewkowycz, J.~Maldacena and S.~J. Suh, \emph{{Relative
  entropy equals bulk relative entropy}},
  \href{https://doi.org/10.1007/JHEP06(2016)004}{\emph{JHEP} {\bfseries 06}
  (2016) 004} [\href{https://arxiv.org/abs/1512.06431}{{\ttfamily
  1512.06431}}].

\bibitem{Almheiri:2014lwa}
A.~Almheiri, X.~Dong and D.~Harlow, \emph{{Bulk Locality and Quantum Error
  Correction in AdS/CFT}},
  \href{https://doi.org/10.1007/JHEP04(2015)163}{\emph{JHEP} {\bfseries 04}
  (2015) 163} [\href{https://arxiv.org/abs/1411.7041}{{\ttfamily 1411.7041}}].

\bibitem{Dong:2016eik}
X.~Dong, D.~Harlow and A.~C. Wall, \emph{{Reconstruction of Bulk Operators
  within the Entanglement Wedge in Gauge-Gravity Duality}},
  \href{https://doi.org/10.1103/PhysRevLett.117.021601}{\emph{Phys. Rev. Lett.}
  {\bfseries 117} (2016) 021601}
  [\href{https://arxiv.org/abs/1601.05416}{{\ttfamily 1601.05416}}].

\bibitem{Faulkner:2017vdd}
T.~Faulkner and A.~Lewkowycz, \emph{{Bulk locality from modular flow}},
  \href{https://arxiv.org/abs/1704.05464}{{\ttfamily 1704.05464}}.

\bibitem{Harlow:2016vwg}
D.~Harlow, \emph{{The Ryu-Takayanagi Formula from Quantum Error Correction}},
  \href{https://arxiv.org/abs/1607.03901}{{\ttfamily 1607.03901}}.

\bibitem{Dong:2016fnf}
X.~Dong, \emph{{The Gravity Dual of Renyi Entropy}},
  \href{https://doi.org/10.1038/ncomms12472}{\emph{Nature Commun.} {\bfseries
  7} (2016) 12472} [\href{https://arxiv.org/abs/1601.06788}{{\ttfamily
  1601.06788}}].

\bibitem{Iyer:1994ys}
V.~Iyer and R.~M. Wald, \emph{{Some properties of Noether charge and a proposal
  for dynamical black hole entropy}},
  \href{https://doi.org/10.1103/PhysRevD.50.846}{\emph{Phys.Rev.} {\bfseries
  D50} (1994) 846} [\href{https://arxiv.org/abs/gr-qc/9403028}{{\ttfamily
  gr-qc/9403028}}].

\bibitem{Fursaev:2013fta}
D.~V. Fursaev, A.~Patrushev and S.~N. Solodukhin, \emph{{Distributional
  Geometry of Squashed Cones}},
  \href{https://arxiv.org/abs/1306.4000}{{\ttfamily 1306.4000}}.

\bibitem{Miao:2014nxa}
R.-X. Miao and W.-z. Guo, \emph{{Holographic Entanglement Entropy for the Most
  General Higher Derivative Gravity}},
  \href{https://doi.org/10.1007/JHEP08(2015)031}{\emph{JHEP} {\bfseries 08}
  (2015) 031} [\href{https://arxiv.org/abs/1411.5579}{{\ttfamily 1411.5579}}].

\bibitem{Miao:2015iba}
R.-X. Miao, \emph{{Universal Terms of Entanglement Entropy for 6d CFTs}},
  \href{https://doi.org/10.1007/JHEP10(2015)049}{\emph{JHEP} {\bfseries 10}
  (2015) 049} [\href{https://arxiv.org/abs/1503.05538}{{\ttfamily
  1503.05538}}].

\bibitem{Camps:2016gfs}
J.~Camps, \emph{{Gravity duals of boundary cones}},
  \href{https://doi.org/10.1007/JHEP09(2016)139}{\emph{JHEP} {\bfseries 09}
  (2016) 139} [\href{https://arxiv.org/abs/1605.08588}{{\ttfamily
  1605.08588}}].

\bibitem{Castro:2014tta}
A.~Castro, S.~Detournay, N.~Iqbal and E.~Perlmutter, \emph{{Holographic
  entanglement entropy and gravitational anomalies}},
  \href{https://doi.org/10.1007/JHEP07(2014)114}{\emph{JHEP} {\bfseries 07}
  (2014) 114} [\href{https://arxiv.org/abs/1405.2792}{{\ttfamily 1405.2792}}].

\bibitem{Rosenhaus:2014woa}
V.~Rosenhaus and M.~Smolkin, \emph{{Entanglement Entropy: A Perturbative
  Calculation}}, \href{https://doi.org/10.1007/JHEP12(2014)179}{\emph{JHEP}
  {\bfseries 12} (2014) 179} [\href{https://arxiv.org/abs/1403.3733}{{\ttfamily
  1403.3733}}].

\bibitem{Lewkowycz:2014jia}
A.~Lewkowycz and E.~Perlmutter, \emph{{Universality in the geometric dependence
  of Renyi entropy}},
  \href{https://doi.org/10.1007/JHEP01(2015)080}{\emph{JHEP} {\bfseries 01}
  (2015) 080} [\href{https://arxiv.org/abs/1407.8171}{{\ttfamily 1407.8171}}].

\bibitem{Iyer:1995kg}
V.~Iyer and R.~M. Wald, \emph{{A Comparison of Noether charge and Euclidean
  methods for computing the entropy of stationary black holes}},
  \href{https://doi.org/10.1103/PhysRevD.52.4430}{\emph{Phys.Rev.} {\bfseries
  D52} (1995) 4430} [\href{https://arxiv.org/abs/gr-qc/9503052}{{\ttfamily
  gr-qc/9503052}}].

\bibitem{Jafferis:2014lza}
D.~L. Jafferis and S.~J. Suh, \emph{{The Gravity Duals of Modular
  Hamiltonians}},  \href{https://arxiv.org/abs/1412.8465}{{\ttfamily
  1412.8465}}.

\bibitem{Donnelly:2014fua}
W.~Donnelly and A.~C. Wall, \emph{{Entanglement entropy of electromagnetic edge
  modes}}, \href{https://doi.org/10.1103/PhysRevLett.114.111603}{\emph{Phys.
  Rev. Lett.} {\bfseries 114} (2015) 111603}
  [\href{https://arxiv.org/abs/1412.1895}{{\ttfamily 1412.1895}}].

\bibitem{Donnelly:2015hxa}
W.~Donnelly and A.~C. Wall, \emph{{Geometric entropy and edge modes of the
  electromagnetic field}},
  \href{https://doi.org/10.1103/PhysRevD.94.104053}{\emph{Phys. Rev.}
  {\bfseries D94} (2016) 104053}
  [\href{https://arxiv.org/abs/1506.05792}{{\ttfamily 1506.05792}}].

\bibitem{Casini:2013rba}
H.~Casini, M.~Huerta and J.~A. Rosabal, \emph{{Remarks on entanglement entropy
  for gauge fields}},
  \href{https://doi.org/10.1103/PhysRevD.89.085012}{\emph{Phys. Rev.}
  {\bfseries D89} (2014) 085012}
  [\href{https://arxiv.org/abs/1312.1183}{{\ttfamily 1312.1183}}].

\bibitem{Allais:2014ata}
A.~Allais and M.~Mezei, \emph{{Some results on the shape dependence of
  entanglement and Rényi entropies}},
  \href{https://doi.org/10.1103/PhysRevD.91.046002}{\emph{Phys. Rev.}
  {\bfseries D91} (2015) 046002}
  [\href{https://arxiv.org/abs/1407.7249}{{\ttfamily 1407.7249}}].

\bibitem{Faulkner:2015csl}
T.~Faulkner, R.~G. Leigh and O.~Parrikar, \emph{{Shape Dependence of
  Entanglement Entropy in Conformal Field Theories}},
  \href{https://doi.org/10.1007/JHEP04(2016)088}{\emph{JHEP} {\bfseries 04}
  (2016) 088} [\href{https://arxiv.org/abs/1511.05179}{{\ttfamily
  1511.05179}}].

\bibitem{Faulkner:2016mzt}
T.~Faulkner, R.~G. Leigh, O.~Parrikar and H.~Wang, \emph{{Modular Hamiltonians
  for Deformed Half-Spaces and the Averaged Null Energy Condition}},
  \href{https://doi.org/10.1007/JHEP09(2016)038}{\emph{JHEP} {\bfseries 09}
  (2016) 038} [\href{https://arxiv.org/abs/1605.08072}{{\ttfamily
  1605.08072}}].

\bibitem{Marolf:2016dob}
D.~Marolf and A.~C. Wall, \emph{{State-Dependent Divergences in the
  Entanglement Entropy}},
  \href{https://doi.org/10.1007/JHEP10(2016)109}{\emph{JHEP} {\bfseries 10}
  (2016) 109} [\href{https://arxiv.org/abs/1607.01246}{{\ttfamily
  1607.01246}}].

\bibitem{Wald:1993nt}
R.~M. Wald, \emph{{Black hole entropy is the Noether charge}},
  \href{https://doi.org/10.1103/PhysRevD.48.R3427}{\emph{Phys.Rev.} {\bfseries
  D48} (1993) 3427} [\href{https://arxiv.org/abs/gr-qc/9307038}{{\ttfamily
  gr-qc/9307038}}].

\bibitem{Lashkari:2016idm}
N.~Lashkari, J.~Lin, H.~Ooguri, B.~Stoica and M.~Van~Raamsdonk,
  \emph{{Gravitational Positive Energy Theorems from Information
  Inequalities}}, \href{https://doi.org/10.1093/ptep/ptw139}{\emph{PTEP}
  {\bfseries 2016} (2016) 12C109}
  [\href{https://arxiv.org/abs/1605.01075}{{\ttfamily 1605.01075}}].

\bibitem{Faulkner:2013yia}
T.~Faulkner, \emph{{The Entanglement Renyi Entropies of Disjoint Intervals in
  AdS/CFT}},  \href{https://arxiv.org/abs/1303.7221}{{\ttfamily 1303.7221}}.

\bibitem{Camps:2014voa}
J.~Camps and W.~R. Kelly, \emph{{Generalized gravitational entropy without
  replica symmetry}},
  \href{https://doi.org/10.1007/JHEP03(2015)061}{\emph{JHEP} {\bfseries 03}
  (2015) 061} [\href{https://arxiv.org/abs/1412.4093}{{\ttfamily 1412.4093}}].

\bibitem{Faulkner:2014jva}
T.~Faulkner, \emph{{Bulk Emergence and the RG Flow of Entanglement Entropy}},
  \href{https://doi.org/10.1007/JHEP05(2015)033}{\emph{JHEP} {\bfseries 05}
  (2015) 033} [\href{https://arxiv.org/abs/1412.5648}{{\ttfamily 1412.5648}}].

\bibitem{Faulkner:2017tkh}
T.~Faulkner, F.~M. Haehl, E.~Hijano, O.~Parrikar, C.~Rabideau and
  M.~Van~Raamsdonk, \emph{{Nonlinear Gravity from Entanglement in Conformal
  Field Theories}},  \href{https://arxiv.org/abs/1705.03026}{{\ttfamily
  1705.03026}}.

\bibitem{Cotler:2017erl}
J.~Cotler, P.~Hayden, G.~Salton, B.~Swingle and M.~Walter, \emph{{Entanglement
  Wedge Reconstruction via Universal Recovery Channels}},
  \href{https://arxiv.org/abs/1704.05839}{{\ttfamily 1704.05839}}.

\bibitem{Callan:1992rs}
C.~G. Callan, Jr., S.~B. Giddings, J.~A. Harvey and A.~Strominger,
  \emph{{Evanescent black holes}},
  \href{https://doi.org/10.1103/PhysRevD.45.R1005}{\emph{Phys. Rev.} {\bfseries
  D45} (1992) R1005} [\href{https://arxiv.org/abs/hep-th/9111056}{{\ttfamily
  hep-th/9111056}}].

\bibitem{Russo:1992ax}
J.~G. Russo, L.~Susskind and L.~Thorlacius, \emph{{The Endpoint of Hawking
  radiation}}, \href{https://doi.org/10.1103/PhysRevD.46.3444}{\emph{Phys.
  Rev.} {\bfseries D46} (1992) 3444}
  [\href{https://arxiv.org/abs/hep-th/9206070}{{\ttfamily hep-th/9206070}}].

\bibitem{Myers:1994sg}
R.~C. Myers, \emph{{Black hole entropy in two-dimensions}},
  \href{https://doi.org/10.1103/PhysRevD.50.6412}{\emph{Phys. Rev.} {\bfseries
  D50} (1994) 6412} [\href{https://arxiv.org/abs/hep-th/9405162}{{\ttfamily
  hep-th/9405162}}].

\end{thebibliography}\endgroup
\end{document}